\documentclass{aa} 
\usepackage{color} 
\usepackage[squaren, Gray, cdot]{SIunits}
\usepackage{graphicx}
\usepackage[varg]{txfonts}
\usepackage{natbib}
\bibpunct{(}{)}{;}{a}{}{,} % to follow the A&A style
\begin{document}
   \title{A fast tree-based method for estimating column densities\\ in Adaptive Mesh Refinement codes}
% Column density estimation: Tree-based grid method 
%
   \subtitle{Influence of UV radiation field on the structure of molecular clouds.}

   \author{Valeska Valdivia \inst{\ref{inst1},\ref{inst3}}
          \and
          Patrick Hennebelle\inst{\ref{inst1},\ref{inst2}}
	}

	\institute{Laboratoire de radioastronomie, LERMA, Observatoire de Paris, \'Ecole Normale Sup\'erieure (UMR 8112 CNRS), 24 rue Lhomond, 75231 Paris Cedex 05, France\\
	\email{valeska.valdivia@lra.ens.fr}\label{inst1}
	\and %Sorbonne Universités, UPMC Univ Paris06, IFD, 4 Place Jussieu, 75252 PARIS cedex05
	Sorbonne Universit\'es, UPMC Univ Paris06, IFD, 4 place Jussieu, 75252 Paris Cedex 05, France \label{inst3}
	%Universit\'e Pierre et Marie Curie, 4 place Jussieu 75005 Paris, France \label{inst3}
	\and
	Laboratoire AIM, Paris-Saclay, CEA/IRFU/SAp - CNRS - Universit\'e Paris Diderot, 91191 Gif-sur-Yvette Cedex, France\\
	\email{patrick.hennebelle@lra.ens.fr} \label{inst2}
             }

   \date{}

% \abstract{}{}{}{}{} 
% 5 {} token are mandatory
 
  \abstract
   {Ultraviolet radiation plays a crucial role in molecular clouds. Radiation and matter are tightly coupled and their interplay influences the physical and chemical properties of gas. In particular, modeling the radiation propagation requires calculating column densities, which can be numerically expensive in high-resolution multidimensional simulations.  }
   {Developing fast methods for estimating column densities is mandatory if we are interested in the dynamical influence of the radiative transfer. In particular, we focus on the effect of the UV screening on the dynamics and on the statistical properties of molecular clouds.}
   {We have developed a tree-based method for a fast estimate of column densities, implemented in the adaptive mesh refinement code RAMSES. We performed numerical simulations using this method in order to analyze the influence of the screening on the clump formation.}
   {We find that the accuracy for the extinction of the tree-based method is better than $10 \%$, while the relative error for the column density can be much more. We describe the implementation of a method based on precalculating the geometrical terms that noticeably reduces the calculation time. To study the influence of the screening on the statistical properties of molecular clouds we present the probability distribution function (PDF) of gas and the associated temperature per density bin and the mass spectra for different density thresholds.}
   {The tree-based method is fast and accurate enough to be used during numerical simulations since no communication is needed between CPUs when using a fully threaded tree. It is then suitable to parallel computing. We show that the screening for far UV radiation mainly affects the dense gas, thereby favoring low temperatures and affecting the fragmentation. We show that when we include the screening, more  structures are formed with higher densities in comparison to the case that does not include this effect. We interpret this as the result of the shielding effect of dust, which protects the interiors of clumps from the incoming radiation, thus diminishing the temperature and changing locally the Jeans mass. }

   \keywords{molecular clouds --
	ISM --
	column density --
	star formation
               }

\titlerunning{A fast tree-based method for estimating column densities in AMR codes} 
\authorrunning{V. Valdivia \& P. Hennebelle}
   \maketitle

%
%________________________________________________________________

\section{Introduction}

Radiation plays an important role in several astrophysical processes at different scales, and it is dynamically coupled to the behavior of the gas. In particular, the far ultraviolet (FUV) radiation influences the physical and chemical properties of molecular clouds, while the cosmic rays (CRs) are the dominant ionizating source in both diffuse and dense media, controlling the chemistry. The UV radiation is the main heating source for the gas by regulating the molecule formation rates on the grain surface, but at the same time dust grains shield the inner regions of clouds favoring low temperatures. 

The penetration of FUV radiation into molecular clouds has been studied theoretically and numerically \citep{flannery1980, sandell1975, whitworth1975, goicoechea2007} by modeling the properties of dust grains or by simplifying the geometry of the cloud, but crucial for describing the propagation of radiation is to estimate the column densities. Another example is the cosmic-ray ionization rate, which also depends on the value of the column densities. Indeed, several works show that the CR ionization rate decreases with the increasing value of the column density \citep[see][]{takayanagi1973,padovani2009, indriolo2012}.

The correct treatment of the propagation of radiation and the estimation of ionization rates involve calculating column densities, but this calculation can be numerically challenging in high-resolution multidimensional simulations, hence the interest of developing fast and accurate methods. 

Some common approximations include methods based on ray-tracing schemes, such as \cite{razoumov2005}, that define ray domains according to the photon travel direction, or stochastic integration methods, such as \cite{cantalupo2011}, that use a Monte Carlo combined with an adaptive mesh refinement (AMR) strategy for the ray casting. There is a much simpler approach  by \cite{inoue2012}, who assume that the gas is shielded well from UV photons except in the colliding direction and have used a `two-ray' approximation for dealing with the shielding for the UV radiation. But most of these approaches have several disadvantages, and in general they are computationally expensive or not accurate enough when the geometry of the problem is more complicated, as in a turbulent medium. In spite of being intuitive, ray-tracing methods are numerically demanding. For a simulation with $N$ resolution elements (cells or particles), the number of operations is at least on the order of $N^2$ and it requires the exchange of large amounts of data between different CPUs when parallelized. In problems where the gas properties are dynamically affected by gravity and radiation, it is desirable to develop adapted numerical strategies that permit calculation of the radiative transfer \emph{on-the-fly}. Recently some efforts have been made in this direction.  Using the smoothed particle hydrodynamics code GADGET2, \cite{clark2011} introduce a tree-based scheme, called \emph{TreeCol}. This method uses the information already stored in the tree structure of the code to construct a full $4\pi$ sr map of column densities for each element with a gather approach. Since the column densities are calculated while the tree is walked, its computational cost is also on the order of $N \mathrm{log}N$.

\indent In this work we present a simple scheme for estimate column densities that takes advantage of the tree data structure, implemented for AMR codes. Our tree-based method provides a fast and relatively accurate estimation of column densities that can be used in numerical simulations. We have implemented this method in the AMR code RAMSES \citep{teyssier2002}.

%%%%%%%   OUTLINE HERE   %%%%%%%%%%%%%%
In the following section we present the radiative transfer problem. In section \ref{sec_method} we present an overview of the tree-data structure and introduce our tree-based grid method, as well as our implementation. We also describe a strategy for optimization consisting in precalculation of geometrical contributions, where we introduce further approximations. In section \ref{verification} we present two tests for validation. The first test consists in a uniform spherical cloud and the second one corresponds to a turbulent cloud. At the end of this section, we also include a test for validating the other approximations made during the precalculation of the geometrical contributions. In section \ref{application} we present one application of this method to calculate the UV absorption by dust and its consequences on the dynamics of molecular clouds. Section \ref{conclusions} concludes the paper.

%%%%%%%%%%%%%%%%%%%%%%	(1)	%%%%%%%%%%%%%%%%%%%%%%%%%%%%%%%%%%%%%%%
\section{The radiative transfer problem}\label{rtproblem}

The full radiative transfer equation that describes the interplay of radiation and matter reads as
\begin{equation}
\label{radtransfer}
\frac{d I_{\nu}}{ds} = -\alpha_{\nu} I_{\nu} + j_{\nu}. 
\end{equation}
\noindent where $I_{\nu}$ is the specific intensity, $\alpha_{\nu}$ the absorption coefficient, and $j_{\nu}$ the emission coefficient at  frequency $\nu$. Because we are interested in the early evolution of molecular clouds, when stars have not been formed yet, we can consider that there are no local sources. Then the solution of Eq.~(\ref{radtransfer}) for a case where there is only absorption will be
\begin{equation}
\label{radtransfersol}
I_{\nu}(s) = I_{\nu}(s_0)\exp\left(-\int_{s_0}^s \alpha_{\nu}(s')ds'\right) = I_{\nu}(s_0)\ e^{-\tau_{\nu}}
\end{equation}

This simplified version of the solution lets us solve the problem by just calculating the optical depth $\tau_{\nu}$ along several directions. But even with these simplifications for a simulation with $N$ resolution elements, the cost of solving the radiative transfer problem is on the order of $N^{5/3}$. This can be easily done in postprocessing, but if we are interested in the dynamical interaction between radiation and matter, it is necessary to calculate the optical depths at each time step. This is extremely expensive in terms of CPU time for relatively large simulations and not practical in parallel computing, so that approached methods must be used. In this paper we compare our results to those obtained with a ray-tracing method, which uses $50$ rays, performed in a postprocessing step.

%%%%%%%%%%%%%%%%%%%%%%%%%%%%%%%%%%%%%%%%%%%%%%%%%%%%%%%%%%%%%%%%%%%

\section{Tree-based method} \label{sec_method}

Astrophysical fluids can be treated numerically by using either $(i)$ a Lagrangian approach, with either $N-$body or SPH codes \citep{benz_1988,springel_et_al_2001, hubber_et_al_2011}, where nodes follow the material particles, or $(ii)$ an Eulerian approach with patch-based codes \citep{fryxell_et_al_2000, almgren_et_al_2010, mignone_et_al_2012, enzo_2013}, where each element is fixed and describes how material flows through the grids. 

Several of these astrophysical codes are based on a tree-data structure that consists in a  hierarchical structure where the simulation domain is recursively split into smaller units or \emph{nodes}. These nodes can be cut recursively to four (2D) or eight (3D) smaller "daughter" nodes from the largest node, the \emph{root}, which contains the whole simulation, to the \emph{leaves} that do not contain any substucture. Each node knows its parent node and can access all the other nodes by walking the tree. The numerical cost for a simulation with $N$ resolution elements is proportional to $N \mathrm{log} N$ \citep{barneshut1986,barneshut1989}. 

Since most astrophysical problems span wide ranges of spatial scales, to correcly describe them, it is necessary to use resolutions comparable to the smallest scales. Adaptive mesh refinement (AMR) techniques allow the resolution to be adapted in different regions. This method was introduced for the first time in \cite{berger1984} as an adaptative finite difference method for solving partial differential equations using nested grids in a patch-based AMR. The fully threaded tree (FTT) was proposed by \cite{khokhlov_1998}, where the tree is threaded at all levels and the refinement is done on a cell-by-cell basis. % {\color{green}that are treated as elements of the tree}.

The RAMSES code \citep{teyssier2002} is a grid-based solver with AMR. The refinement levels are labeled $\ell$. The coarse level ($\ell =0$) corresponds to the base of the tree data structure and contains the whole simulation box.  The refinement is done recursively on a cell-by-cell
 basis by splitting the cell into $2^3$ daughter cells that constitute an \emph{oct}, the basic elements in the data structure. The cells in an \emph{oct} are indexed by the \emph{oct index}, and this index is distributed around the center of the \emph{oct} given a specific distribution.

This code has been parallelized using the MPI library and it uses locally essential trees \citep{warren1993}, thus all the information is local. Each CPU knows the full tree up to a given resolution, therefore each cell can recursively access the information from the coarser levels.

\subsection{General Idea}
%%%%%%%%%%%%%%%%%	(2)	%%%%%%%%%%%%%%%%%%%%%%%%%%%%%%%%%%%%%%%%%%
The tree-based method for estimating column densities is based on the fact that any distant cell substends a small angle and therefore its contribution to the extinction along the line of sight will be diluted. Then it is suitable to approximate the distant structured cells with cells at lower resolution. 

For each target cell, column densities can be estimated by summing up all the contributions of cells along each line of sight and decreasing resolution with distance. Because all the information can be accessed by walking the tree and knowing the density, the distance, and the size of a cell at a given resolution level, we can calculate the contribution to the column density as the product of the density and the distance covered through the crossed cell along the  line of sight. %(???)
%%%%%%%%%%%%%%%%%%%%%%%%%%%%%%%%%%%%%%%%%%%%%%%%%%%%%%%%%%%%%%%%%%

%%%%%%% IMPLEMENTATION    %%%%%%%%%%%%
\subsection{Implementing and calculating the extinction}

%%%%%%%             intro             %%%%%%%%

Since molecular clouds are embedded in the interstellar radiation field (ISRF), we are particularly interested in the influence of the far-interstellar radiation field triggered by the radiation of UV photons from OB stars. \cite{habing1968} suggested that this radiation density could be constant throughout space. Here we consider an incident UV field,  the Draine field \citep{draine1978}, which is supposed to be monochromatic, that corresponds to $G_0 \approx 1.7$ in Habing units, attenuated by dust \citep{wolfire1995}. 

We have implemented the calculation of column densities in order to estimate the attenuation for the UV radiation from the ISRF, described by the parameter $G_0$ \citep{habing1968}. From Eq.~(\ref{radtransfersol}) we can define the \emph{attenuation factor} $\chi$, calculated as the mean value of the extinction seen by a given cell and calculated as

   \begin{eqnarray}
	\chi &=& \frac{1}{4 \pi} \iint\limits_{4 \pi} \,   \mathrm{e}^{-\tau(\theta , \phi ) }\, \mathrm{d}\Omega\,, \label{eq_chi}\\
	\tau (\theta , \phi ) &=& \int _0^r  K (\theta , \phi ) \mathrm{d} r =     \sigma \mathcal{N} (\theta , \phi ) \label{eq_tau}
   \end{eqnarray} 

\noindent where $\tau$ is the optical depth along the line of sight defined by $\theta$ and $\phi$ in spherical coordinates, $K$ is the extinction coefficient, and $\sigma$ the effective attenuation cross section for the dust grains at $\lambda = 1000 ~\angstrom$. Here we use $\sigma_{\mathrm{d}, 1000} = 2 \times 10^{-21} \mathrm{cm}^2 $ \citep{draine1996}, and $\mathcal{N}$ is the total column density of hydrogen.

%%%%% DISCRETIZATION

\begin{figure} \resizebox{\hsize}{!}
{\includegraphics{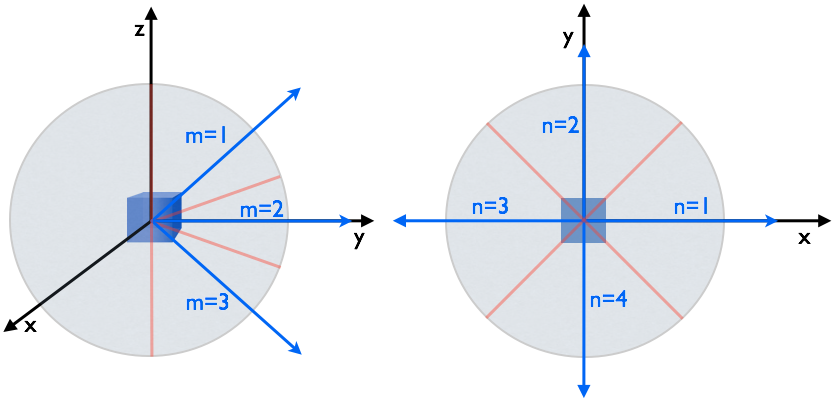}} 
%{\includegraphics{}} 
\caption{Example of the discretization for $N=4$ and $M=3$. The red lines show the bounds of the solid angles, while the blue arrows show the representative directions $(\theta, \phi)$ for the column densities.}
\label{get_mn}
\end{figure}

To calculate the column density maps and build the extinction maps, there are two different possible approaches. The first one uses the \emph{ray-tracing} approach, where the column densities are defined by a single `ray'. The second one uses a \emph{gather} approach, where all the matter that falls in a given solid angle is gathered and added to the column density. The approach that we have adopted is quite similar to the ray-tracing. However, it permits us to add part of the matter that belong to neighboring cells by diminishing the resolution for distant cells, in such a way that the density corresponds to a mean. This has the advantage of considering the contribution to the screening from neighboring cells that could be missed by using a simple ray-tracing algorithm. This approximation represents a lower computational cost than for an exact gather approach, where we would be required to perform more complex calculations.
  
%OR \citet{clark2011}
We call the \emph{target cell} the cell for which we currently want to estimate column densities, in order to keep the same nomenclature as \cite{clark2011}, and we call \emph{treated cells} those contributing to the column density seen by the \emph{target cell}. For the resolution we use the notation
\[
\begin{array}{lp{0.8\linewidth}}
\ell_{0}	& the resolution level of the target cell \\
\ell		& the resolution level required for a cell contributing to the column density ($\ell < \ell_{0}$).
\end{array}
\]
 
For this approach we define directions based on a spherical projection centered on the target cell. We discretize the azimuthal angle in $N$ regular intervals ($\delta\phi = 2 \pi/N$) and the polar angle in $M$ irregular intervals constrained to the $\delta \cos \theta = 2/M$ constant in order to cover equal solid angles. (Fig.~\ref{get_mn} shows how these directions are defined.) Then the directions are labeled by two indices $m$ and $n$, and the representative angles for these directions are given by
\begin{eqnarray}
	\theta_m & = & \arccos\left( 1 + \frac{(1-2 m)}{M}\right) \\
	\label{theta}
	\phi_n & = & \frac{2\pi}{N}(n-1) 
	\label{phi}
\end{eqnarray}

%%%%%%%%% NEW PH %%%%%%%%%%%
\noindent where $m \in [1, M]$ and $n \in [1, N]$. Since solid angles are equal and cells have uniform densities, we can rewrite Eq.~(\ref{eq_chi}) as
\begin{eqnarray}
\label{chi_discret}
\chi &=& \frac{1}{M\times N}\sum_{m}\sum_{n} \mathrm{e}^{-\sigma \mathcal{N}(m,n) }\\
\label{N_discret1}
\mathcal{N}(m,n) &=& \sum_{i} n_i \Delta x_{i} (m, n)
\end{eqnarray}

\noindent where the index $i$ stands for the cell number, $n_i$ and $\Delta x_i (m, n)$ correspond to the number density of the cell $i$ and to the distance crossed through the cell $i$ in direction ($m, n$) with respect to the target cell, respectively.

%%%%%%% 3 regions int loc ext %%%%%%%%
%%%%%%% cubic domains          %%%%%%%%
%Each contribution will be given as the product of the distance covered through the crossed cell (???) and its density at the required resolution. 

%%%%=== INTEGRATION
%%%
%%%For the integration of the colum density along each direction we decrease the resolution with distance. The contribution to the column density can be calculated as the distance crossed inside the emph{treated cell} multiplied by its local density at a given resolution. 
%%%%-------------------------------

%(EXTERNAL INTERNAL ETC)

   \begin{figure}
   \centering
   \includegraphics[width=5cm]{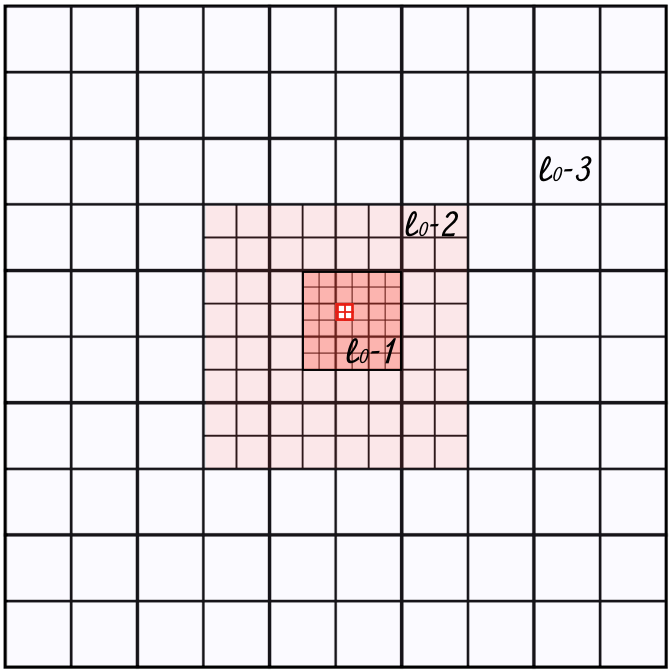} %Pres_LERMA24}
%   \includegraphics[width=5cm]{} %Pres_LERMA24}
%%%Construction of concentric cubic domains around the grid that contains the target cell (in red). 
   \caption{ Central region containing the target cell and the cells that belong to its oct (sibling cells) at the resolution level of the target cell $\ell_0$. The subsequent domains are constructed around the grid decreasing the resolution consecutively. In this manner the $i$-th shell will contain cells at level $\ell_0 -i$.   
}
    \label{Fig_cubicdomain}
    \end{figure}

For integrating column densities $\mathcal{N}(m,n)$ along the different directions, we define two main regions. The first region contains the cell itself and its siblings (the cells that belong to the same oct, sharing a common parent node), and the contribution to the column density is calculated at resolution level $\ell_0$. The outer region contains the rest of the cells in the simulation box, and the contribution to the column density is estimated by decreasing the resolution with the distance. For the outer region, we construct cubic concentric domains, where cells are treated at the same resolution. For each level of resolution, starting from $\ell= \ell_0 -1$,  we define a cubic shell as
\begin{itemize}
	\item We set the new center as the center of the grid that contains the target cell at level $\ell+1$ (or equivalently the center of the unrefined cell at level $\ell$); 
	\item We set the inner limit as the external limit of the previous shell;
	\item We define the outer limits for each cartesian direction by taking two or three neighboring cells into account at level  $\ell$ in order to fully cover their parent grid.
\end{itemize}
The procedure is repeated recursively up to the border of the box. These cubic concentric shells define at which resolution the cells are taken into account for the calculation of the column density. Figure~\ref{Fig_cubicdomain} illustrates the construction of these concentric cubic domains. 

%%%%%%% add contrib m,n        %%%%%%%%
To integrate the column densities in each direction we define three contributions: the \emph{internal} contribution, corresponding to the \emph{target cell} itself, the \emph{local} contribution, given by the sibling cells, and the \emph{external} contribution, given by the cells in the outer region.

For the \emph{internal} and \emph{local} contributions, directions are defined with respect to the center of the \emph{target cell}. On the other hand, for the \emph{external} contribution, directions are defined with respect to the center of the grid that contains the \emph{target cell} at level $\ell_0$. This \emph{external} contribution will be the same for all the sibling cells that belong to the same oct. 
Using Eq.~(\ref{N_discret1}) and rewriting the sum as a sum over concentric shells $C_\ell$, we can use the fact that all the cells that belong to the same shell have the same size $dx_\ell$ to write the external contribution as 

\begin{eqnarray}
\label{N_discret0}
\mathcal{N}_{ext}(m,n) &=&  \sum_{\ell} \sum_{i \in C_\ell} n_i \Delta x_i (m, n, \ell) \\
\label{deltax}
\Delta x_i (m, n, \ell) &=&  dx_\ell \times \mathcal{K}_i(m,n,\ell)
%=  \sum_{\ell} \sum_{i\ \in \ C_\ell}  n_i dx_\ell \times \mathcal{K}(\mathbf{r}_i - \mathbf{r}_{target}, m, n, \ell -\ell_0) }
\end{eqnarray}

\noindent where $C_\ell$ stands for the shell at level $\ell$. The size of the cell at level $\ell$ is given by $dx_\ell = 0.5^\ell L$, with $L$ the size of the simulation box. The multiplicative factor $\mathcal{K}_i$ contains the geometrical corrections to the distance crossed through the cell $i$ with respect to the target cell in the direction $(m, n)$ at level $\ell$. Finally, the column densities along each line of sight can be estimated by adding all the contributions.

%%%%%%% optimization             %%%%%%%%
\subsection{Optimization and precalculation module}\label{optim}

   \begin{figure}
   \centering
   \includegraphics[height=3.5cm]{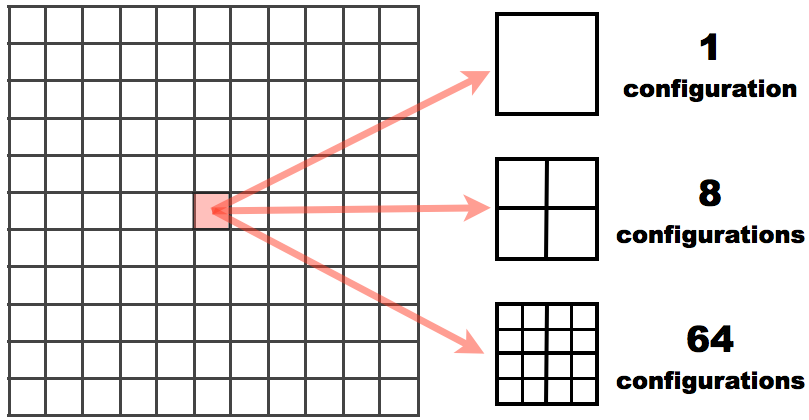}
   \caption{Standard cube showing all the possible configurations for the two-level approximation. The size of each cell in the standard cube is considered to be one unit in terms of the local $dx$.}   %%%FAIRE MIEUX
    \label{stdcube}
    \end{figure}

%%%%%%%%% Verification %%%%%%%%%%%%%
   \begin{figure}
   \centering
   \includegraphics[height=3cm]{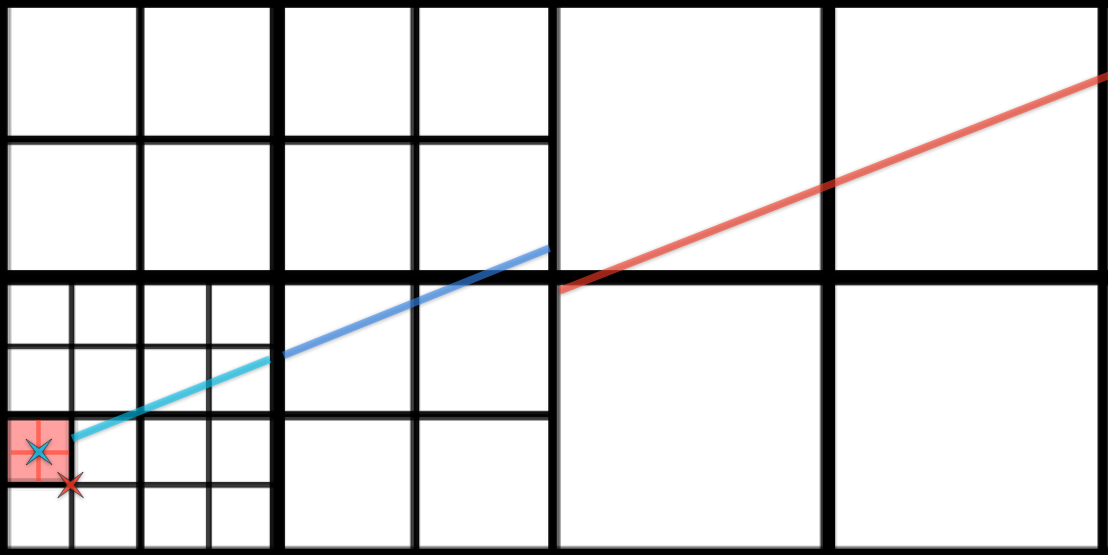}
   \caption{Effect of the two level approximation on estimation of single column densities. The parent grid that contains the target cell is marked in red.}
    \label{2levelcorrection}
    \end{figure}
%%%%%%%%%%%%%%%%%%%%%%%%%%%%%%%

The geometrical factor $\mathcal{K}_i(m,n,\ell)$ in Eq.~(\ref{deltax}) is quite expensive to calculate; however, because the oct structure is self-similar this factor can in principle be calculated and stored at the beginning of the simulation. Then, using the fact that octs are self-similar at different resolution levels and knowing by construction that the outer limit for any cubic shell is at most five times the local $dx$ in each cartesian direction ($\pm x$, $\pm y$, $\pm z$), we can define a standard cube of size $11^3$ total cells (Fig.~\ref{stdcube}) of a one unit size. The cell in the center contains the \emph{target cell}, so we can represent all the possible cases by defining its internal configuration, depending on the difference between the target cell level and the shell level, given by $\Delta \ell = \ell_0 - \ell$. Therefore the total number of configurations is given by $\sum\nolimits_{\Delta \ell = 0} ^{\ell_{max}-2} (2^3)^{\Delta \ell}$. For each one of these cases, we need to calculate the correction for any cell in the cubic shell in any direction. This implies $M\times N \times 11^3$ corrections for each case. This leads to a huge matrix that cannot be stored. We can, however, reduce the size of the matrix by restricting the number of cases considered. This is done by requiring that the target center is shifted for high values of $\Delta \ell$. 

As a first approximation for a shell at level $\ell$, the center of the target cell can be approximated by the center of the grid that contains the target cell at level $\ell +1$. This is equivalent to replacing it by the target cell center at level $\ell$. This is the exact configuration when $\ell_0 - \ell = 1$, but for other levels, the position of reference can be considerably drifted away, and then more precise corrections are needed. We introduce two correction levels in order to take the configuration of the target cell into account with respect to its coarser octs. The first correction level is used when $\ell_0 - \ell = 2$ and it considers the index of the target cell in its oct at level $\ell-1$, so that there are eight possible positions. The second correction level is used for the case $\ell_0 -\ell \geq 3$, where we consider two oct indices. The first one is the oct index of the cell that contains the target cell at level $\ell+2$, and the second one corresponds to the oct index at level $\ell + 3$, generating $64$ possible configurations. This leads to a total of $73$ possible configurations. The effect of this approximation on the estimation of individual column densities is depicted in Fig.~\ref{2levelcorrection}. Then for each configuration we calculate the distance crossed through every cell in the general cube for each one of the directions ($m$, $n$) in order to obtain the corrective factor $\mathcal{K}$ defined in Eq.~(\ref{N_discret}). Finally the pre-calculation module generates a matrix that contains the corrective factors $\mathcal{K}$ for all the configurations considered and a boolean matrix that permits us to know if  the geometrical correction is non-zero. These matrices are calculated just once at the beginning of the simulation.

The external contribution will be calculated as

\begin{equation}
\label{N_discret}
\mathcal{N}_{ext}(m,n) =  \sum_{\ell} dx_\ell \sum_{i\ \in \ C_\ell}  n_i  \times \mathcal{K}(\mathbf{r}_i - \mathbf{r}_{target}, m, n, \Delta \ell),
\end{equation}

\noindent then the corrective factors $\mathcal{K}$ can be found by knowing the indices of the required line of sight $m$ and $n$, the configuration within the octs given by the $\Delta \ell = \ell_0 -\ell$, and the relative position ($\mathbf{r}_i~-~\mathbf{r}_{target}$) of the cells in the oct with respect to the target cell in $dx_\ell$ units. As described later in Sect.~\ref{validationprec}, this speeds up the code by a fair amount without reducing too much the accuracy.

\section{Verification of the method}\label{verification}

In this section we analyze the reliability of our tree-based method by comparing our estimations to a reference. At the end of this section, we analyze the influence of the precalculation of geometrical terms and the approximation introduced in section \ref{optim} on the performance and accuracy. 

To validate of our method we considered two test cases. The first one consisted of a uniform spherical cloud, and the second one corresponded to a turbulent cloud. For both cases we present the total column density maps integrated along each one of the main axes $x$, $y$ and $z$ with respect to the midplanes and the extinction maps as seen by the cells in the midplanes calculated with our tree-based method for the following cases: $6$, $12$, $40$, and $84$ directions. To show the error dependence on the position angle, we present column density projections onto $4\pi \ \mathrm{sr}$ maps as seen by a cell for both test cases. 

The reference maps were calculated in postprocessing with two different methods. For the column density and extinction maps in a plane, we used a ray-tracing approach that uses $50$ rays and takes all the cells of the simulation into account at the highest resolution. It calculates the exact contribution to the column density for each direction using all the cells that are intersected by the ray. For the $4\pi \ \mathrm{sr}$ maps and for the mean column density maps, the reference was calculated with a gather approach. This method divides the computational domain in angular and radial bins. It finds the closest bin for each cell and calculates the fraction of the mass that falls in this bin and in angular neighboring bins. Because the geometrical corrections for cells are much more complicated than for spherical particles, the fraction was calculated assuming that the cell is seen perpendicular to one of its faces. The column density is calculated as the sum of the mean density in the bin multiplied by the radial thickness of the bin. 

These two methods have several inconveniences. They require the information already stored in each CPU, so they are not suitable for parallel computing, and they are extremely expensive in terms of CPU time. In particular, the ray-tracing method has a numerical cost of about $N^{4/3}$ times the number of rays, while the gather method has a numerical cost of about $N^2$, since for each cell all the cells are used for calculating the column densities. The reference maps for the extinction, column density, and mean column density are calculated at the same time. The cost of producing the maps for three slices using the ray-tracing method for a maximum resolution level $\ell = 9$ is $90$ hr of CPU-time, while for the maps done using the gather method, the cost for only one slice is more than $6800$ hr of CPU-time. The enormous cost of the gather approach meant that for the mean column density, we performed only one map at the maximum resolution ($\ell = 9$) for the cut at  $z = 25\ \mathrm{pc}$. For the other cuts, we used only $\ell = 8$.

For all the comparisons, the fractional error is defined by Eq.~(\ref{eq1}). Because we were interested in how the UV field is shielded, the extinction maps were normalized to one, and the difference maps were calculated by Eq.~(\ref{eq2}):

\begin{eqnarray}
	\Delta \mathcal{N} _i &=& \frac{|\mathcal{N}_i-\bar{\mathcal{N} _i}|}{\bar{\mathcal{N} _i}}  \label{eq1}	\\
	\Delta \chi_i &=& |\chi_i-\bar \chi_i|  \label{eq2}
\end{eqnarray}

\noindent where $\bar{\mathcal{N}}$ and $\bar \chi$ stand for the reference maps for the column density and extinction, respectively.

%%%%%%%%% TEST SPHERE  Exact vs Approx %%%%%

\subsection{Spherical uniform cloud}

   \begin{figure}
   \centering
    \includegraphics[width=9cm]{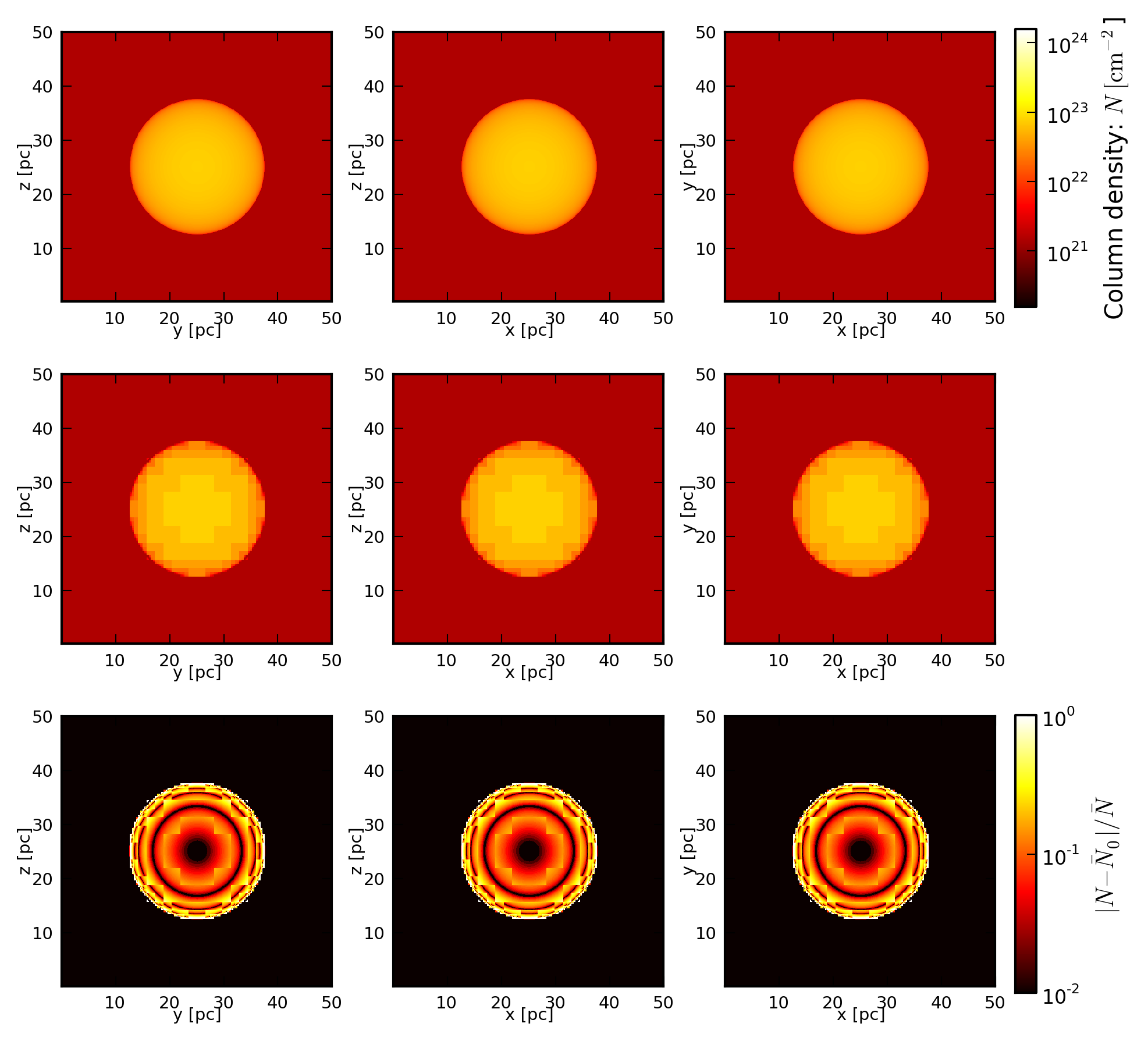}
   \caption{Column density maps integrated along the x, y and z axis for the reference (on the top) and for the tree-based method (center) calculated with respect to the mid planes. The map at the bottom corresponds to the fractional error calculated according to Eq.~(\ref{eq1}).  }
   \label{Fig_sphereN}
   \end{figure}

   \begin{figure}
   \centering
   \includegraphics[width=8cm]{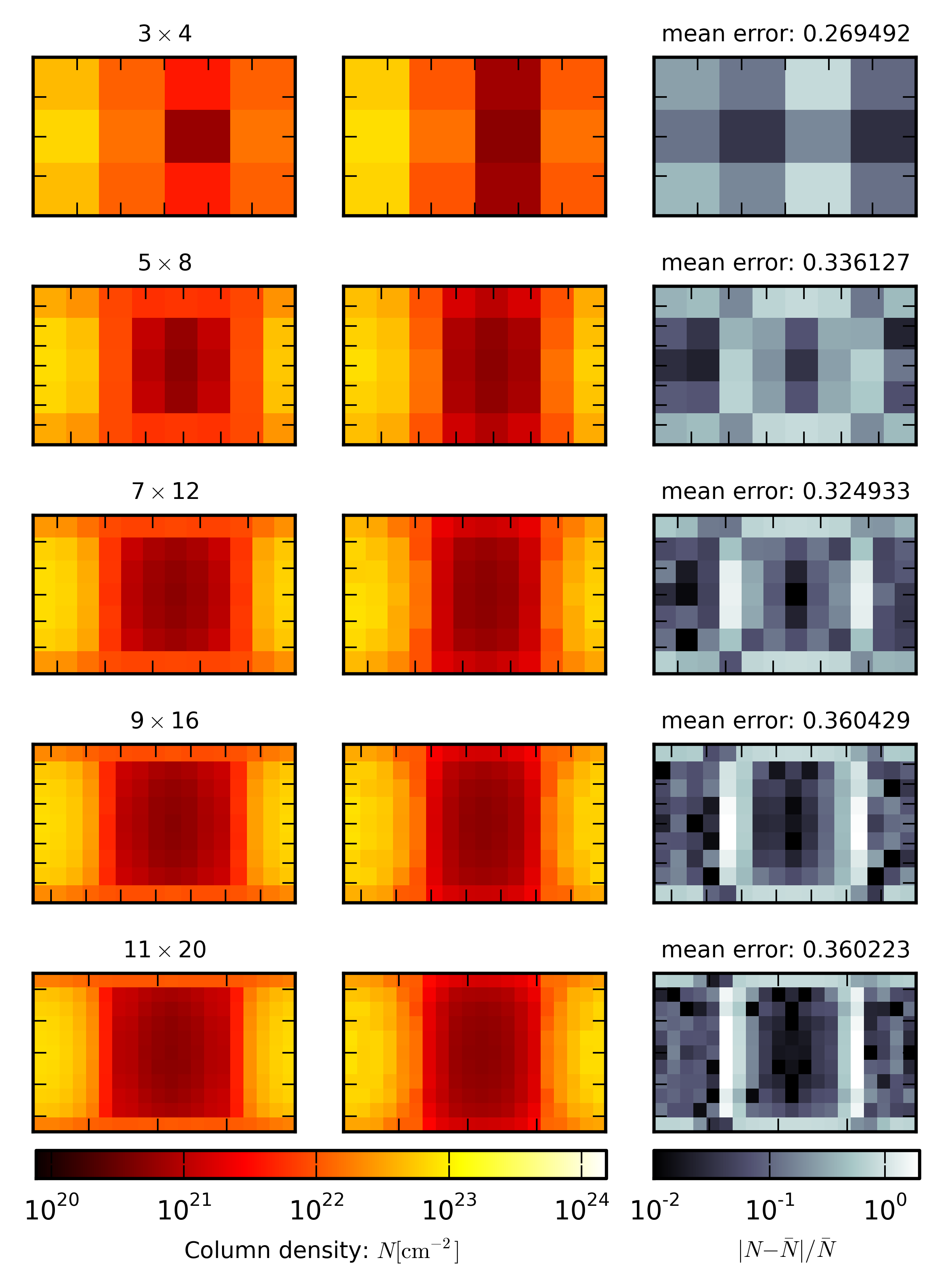}
   \caption{Column densities for the $4\pi \ \mathrm{sr}$ as seen by a cell sitting on the edge of the sphere $(12.5, 25, 25)\  \mathrm{[pc]} )$. The abscissa corresponds to the angle $\phi$ (from $0$ to $2\pi$), and the $\cos \theta$ (from $-1$ to $1$) is given in the ordinate. The reference maps (on the left) have been calculated using the gather approach. The panels at the center present the maps calculated using the tree-based method, while on the right we present the error maps calculated according to Eq.~(\ref{eq1}). From top to bottom: using $12$, $40$, $84$, $144$, and $220$ directions. 
}
   \label{Fig_sph_theta_phi}
   \end{figure}

%PAPER_SPHERE_MN_CORR_L9_TESThot

To test our method we first considered the simple case of a spherical cloud of uniform density. The radius of the spherical cloud is $12.5$~pc, and its number density $1000$ cm$^{-3}$. The cloud is located at the center of a $50$~pc cube with a number density of $10$ cm$^{-3}$. We considered two AMR levels with a maximum resolution equivalent to  $512^{3}$ cells.  The refinement criterion is the density set in such a way that the sphere is refined. 

    \begin{figure}
   \centering
   \includegraphics[width=9.3cm]{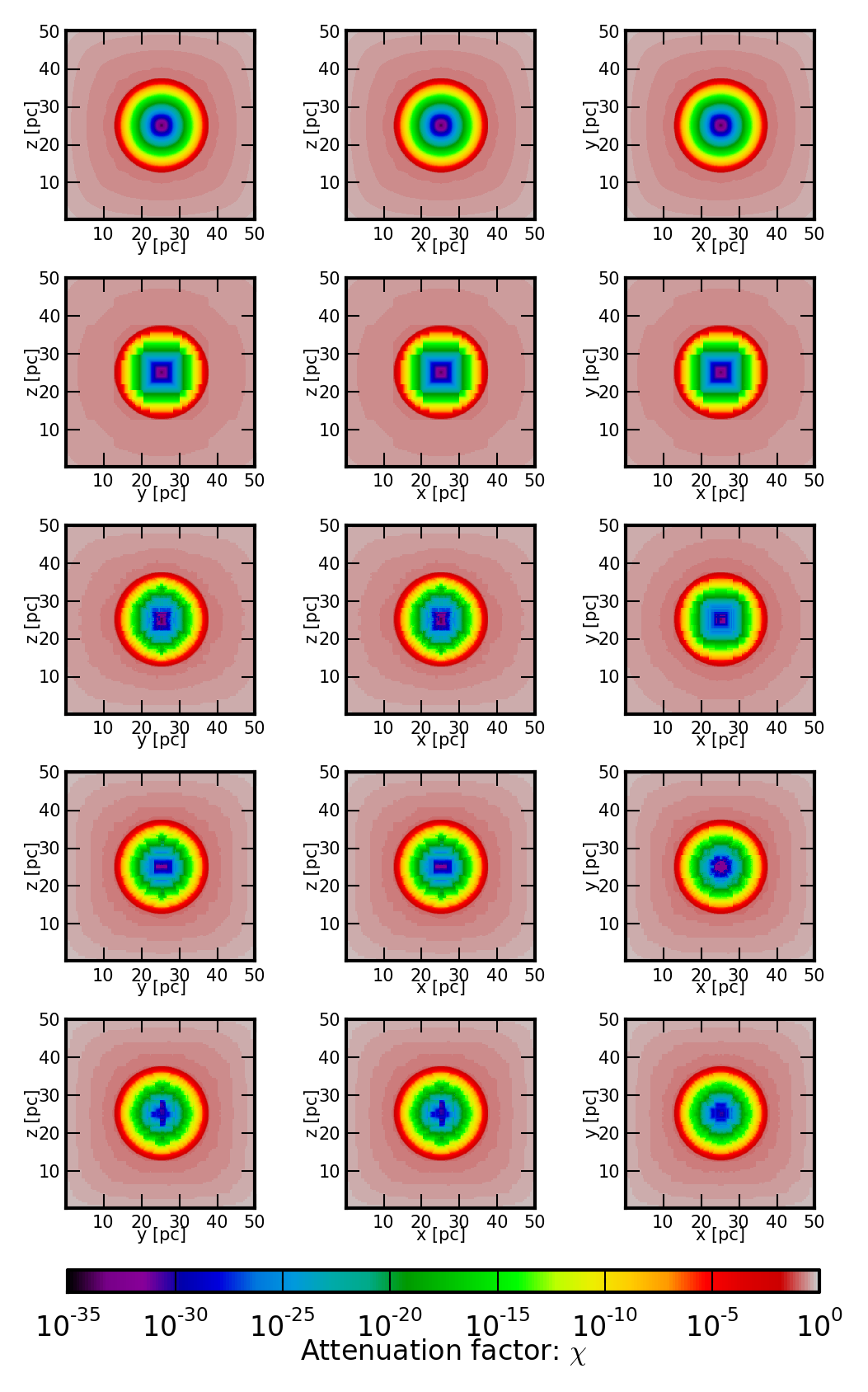}
   \caption{Extinction maps as seen by the cells in the midplanes for $x$, $y$, and $z$. From top to bottom: the reference calculated with a ray-tracing method, then using the tree-based method for $6$, $12$, $40$, and $84$ directions.  }
   \label{Fig_sphereX}
   \end{figure}

   \begin{figure}
   \centering
  \includegraphics[width=9cm]{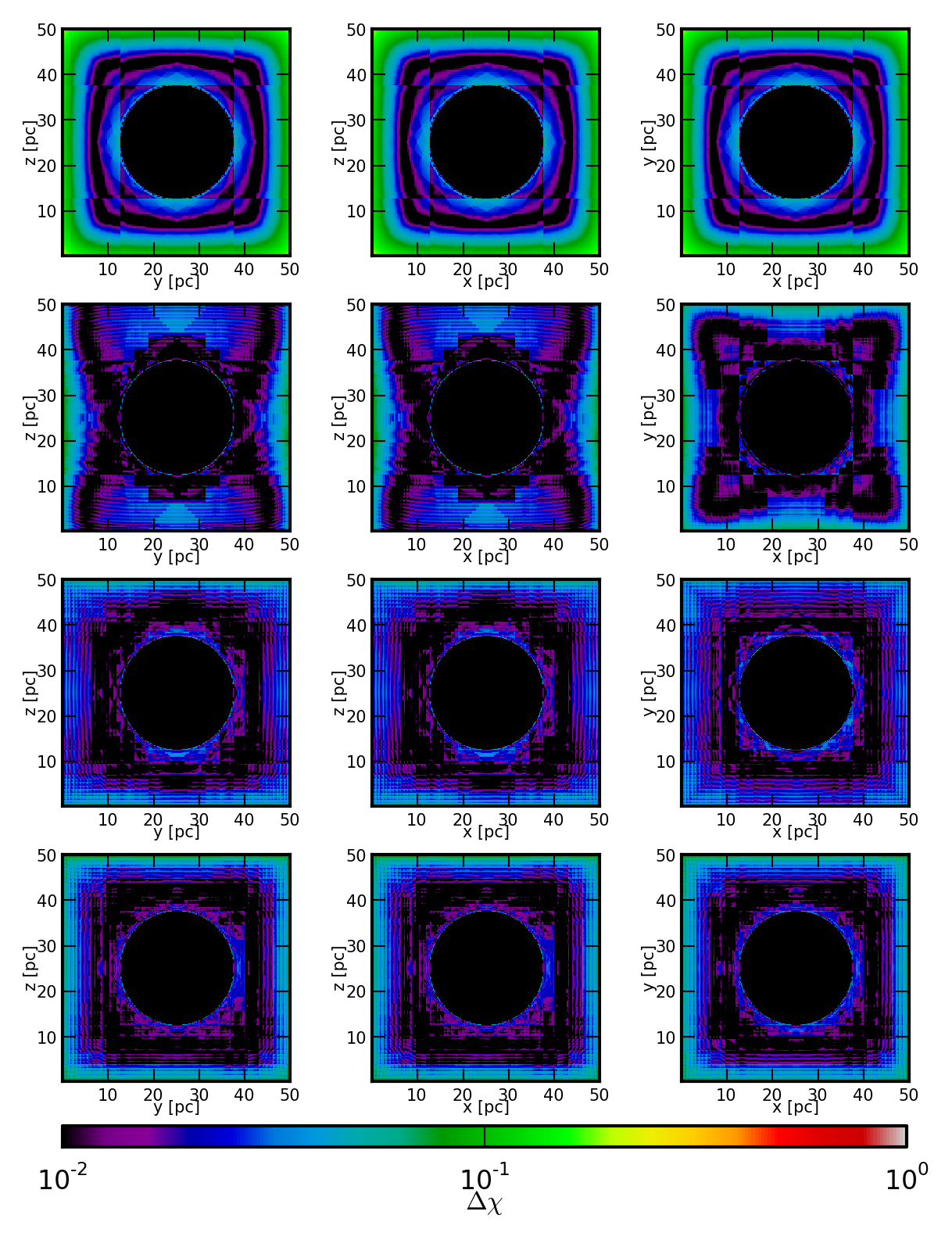} 
  \caption{Difference maps for the extinction as seen by cells in the midplanes for $x$, $y$, and $z$. From top to bottom:  using the tree-based method for $6$, $12$, $40$, and $84$ directions.  }
   \label{err_sphereX}
   \end{figure}

The column density maps integrated along the main axes (in Fig.~\ref{Fig_sphereN}) show that our approximation can reproduce the main features, and it is able to capture the discontinuity in density at the edge of the cloud.  The error maps show that in general the error is lower than $10$\%, but at the edges, the error increases up to more than $100 \%$ in a narrow region. Since most of the volume around the sphere is filled with constant density gas, we calculated the mean error on the sphere alone. This mean error is about $15.2$\% when averaging for the three maps.

%%%%%%%%%%%%%%%%%%%%%%%%%%%%%%%
Since the error in the column density maps is higher at the edge of the sphere, we calculated the column densities as seen by a cell sitting at $(12.25, 0.5, 0.5)\ \mathrm{pc}$. In Fig.~\ref{Fig_sph_theta_phi} we show the column density projection in every direction as seen by a cell at the border of the sphere for $12$, $40$, $84$, $144$, and $220$ directions, using the gather approach and the tree-based method. In the same figure, we present the relative error. For this cell, the center of the sphere is in direction $(\theta, \phi) = (\pi/2, 0)$ so that one half of the sphere is seen at the beginning, and as the angle $\phi$  increases, the column density seen by the cell decreases. When $\phi = \pi$, the direction points outward, where the density is weaker. As the angle $\phi$ keeps increasing, the direction approaches the sphere again. This figure shows that the highest error happens at the edge of the sphere. A similar effect can be seen in Fig. 7 for TreeCol \citep{clark2011}. In our case this error probably occurs because distant cells are not as well described by the tree-based method. 

%%%%%%%%%%%%%%%%%%%%%%%%%%%%%%%

For the extinction, we present a comparison in Fig.~\ref{Fig_sphereX} of the extinction maps for the cells in the midplanes, as defined above. The reference maps calculated with the ray-tracing method are shown in the top row. Then from top to bottom, we present the extinction maps calculated with the tree-based method for $6$, $12$, $40$, and $84$ directions. In Fig.~\ref{err_sphereX} we present the respective difference maps calculated according to Eq.~(\ref{eq2}). For the case with six directions, the mean difference is about $0.036$, while for the cases at $12$, $40$, and $84$ directions the mean difference is less than $0.02$, but the increasing number of directions does not improve the accuracy considerably (Table \ref{table2}).
This is probably because the resolution is getting coarser as we cross cells located farther away. While the number of directions is increasing, the angular resolution remains constant owing to limitations inherent in the method. In this manner, more intervals oversample the region without increasing the angular resolution.

%%%%%%%%% TEST  CLOUD Exact vs Approx %%%%%

\subsection{Turbulent cloud test}

\begin{table}
\caption{Mean difference relative to the reference maps for the extinction.}
\label{table2} % is used to refer this table in the text
\centering
\begin{tabular}{l l l l l }
%%%%%%%%%%%%%%%%%
\hline\hline
Directions			& \multicolumn{4}{c}{$\Delta \chi$}	\\ % table heading
 $M\times N$		& \multicolumn{2}{c}{spherical cloud}		&  \multicolumn{2}{c}{turbulent cloud} 		\\
				&	\textit{mean}	&	\textit{max}	&	\textit{mean}	&	\textit{max}	\\
\hline \\
$6$				&	$0.036$	&	$0.203$	&	$0.079$	&	$0.280$	\\
%\hline
$3 \times 4 = 12$	&	$0.017$	&	$0.094$	&	$0.054$	&	$0.204$	\\
%\hline
$5 \times 8 = 40$	&	$0.014$	&	$0.068$	&	$0.048$	&	$0.164$	\\
%\hline
$7 \times 12= 84$	&	$0.018$	&	$0.090$	&	$0.047$	&	$0.145$	\\
\hline
%%%%%%%%%%%%%%%%%%
\end{tabular}

\end{table}

   \begin{figure}
   \centering
   \includegraphics[width=9cm]{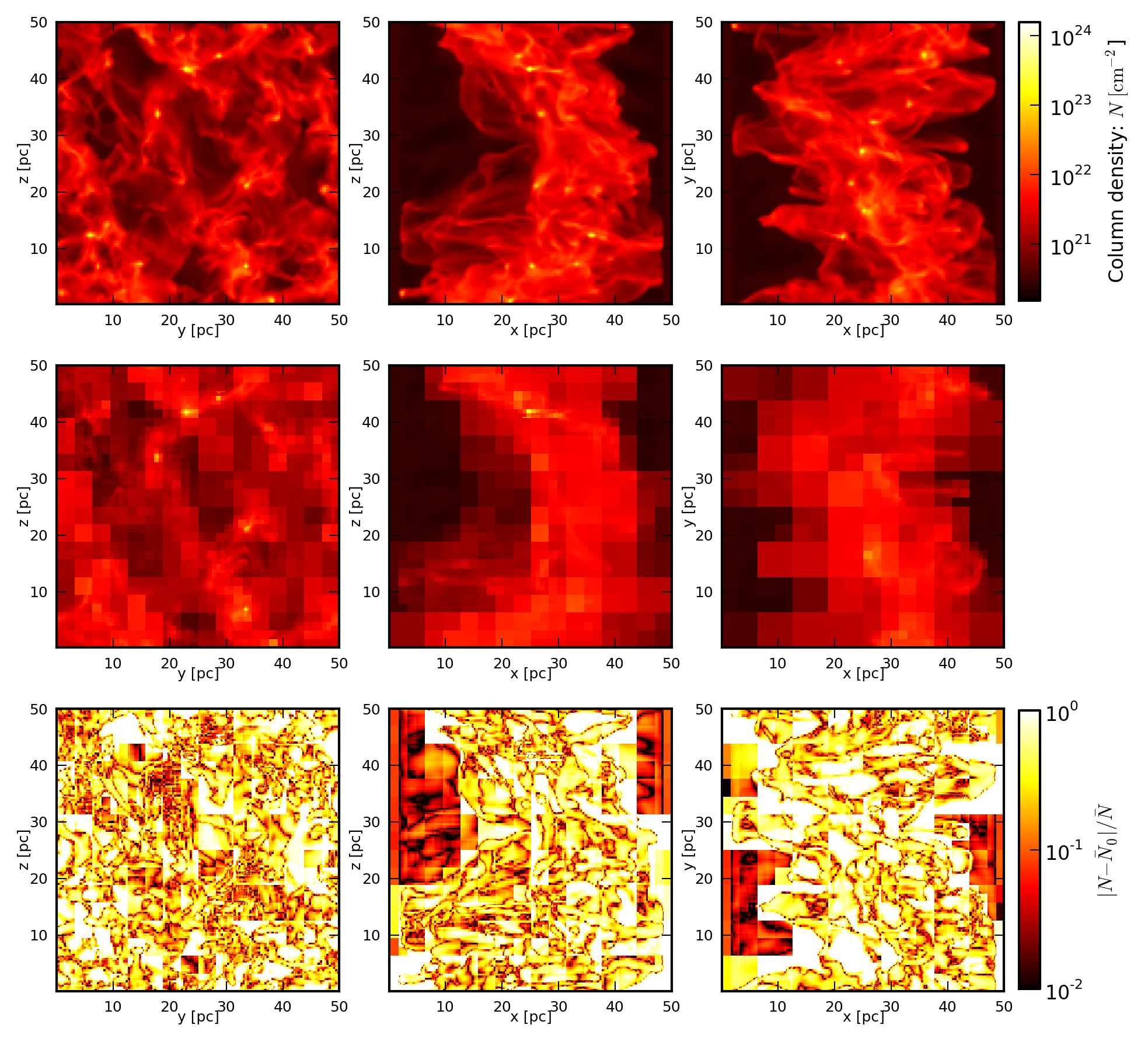}
   \caption{Column density maps for a highly structured cloud, integrated along the $x$, $y$, and $z$ axes (from left to right): using the ray-tracing method (on the top), our tree-based method (middle), and the relative error (on the bottom).}
   \label{Fig_cloudN}
   \end{figure}

   \begin{figure}
   \centering
   \includegraphics[width=8cm]{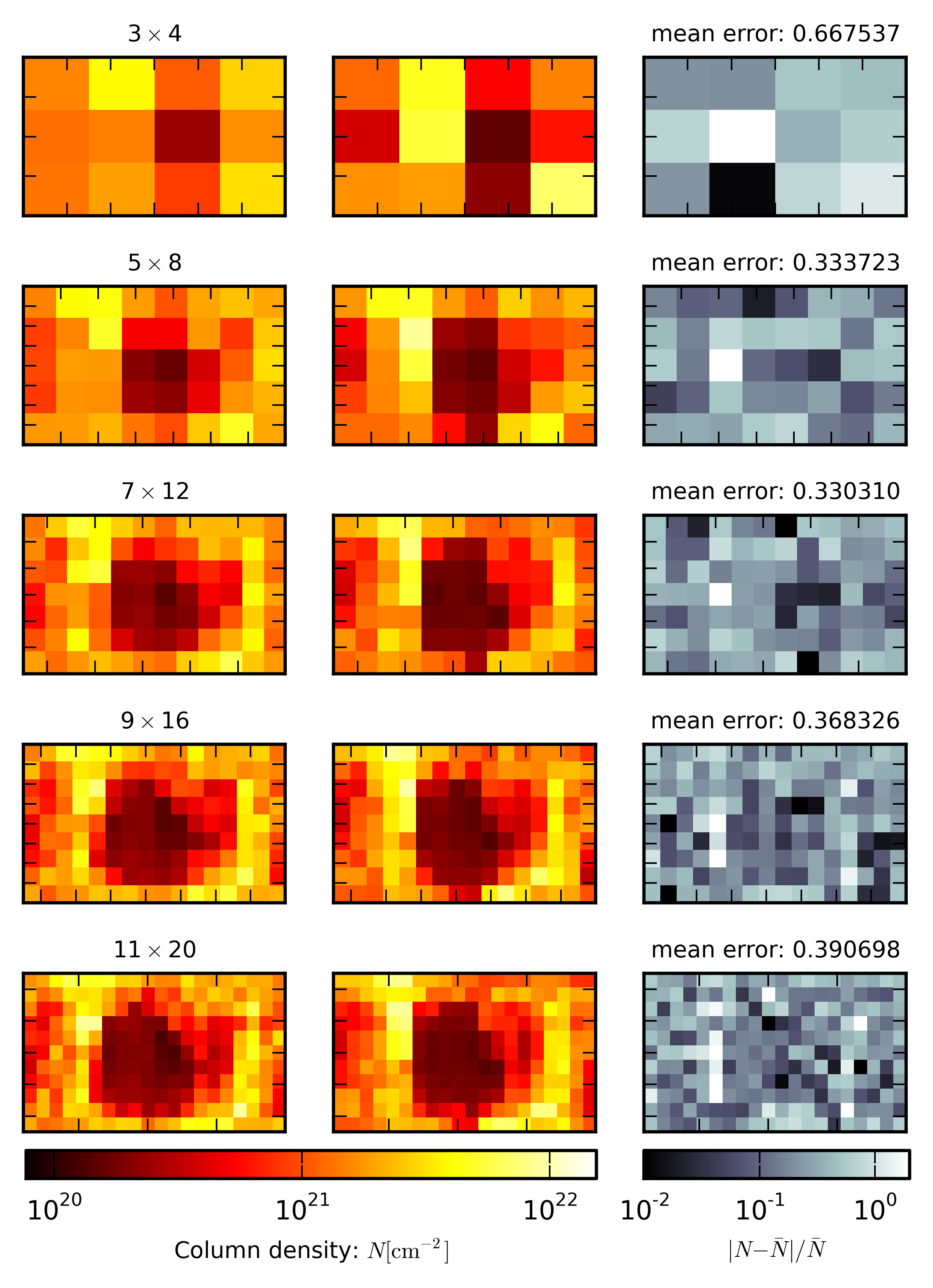}
   \caption{Column densities for the $4\pi \ \mathrm{sr}$ as seen by a cell sitting at the center of the turbulent cloud. As in Fig. \ref{Fig_sph_theta_phi}, the horizontal axis is the azimuthal angle $\phi$ (from $0$ to $2\pi$), and the vertical axis corresponds to $\cos \theta$. On the left we present the reference maps calculated using the gather approach. The panels in the center present the maps calculated using the tree-based method, and on the right we present the error maps calculated according to Eq.~(\ref{eq1}). From top to bottom: using $12$, $40$, $84$, $144$, and $220$ directions. 
}
   \label{Fig_cloud_theta_phi}
   \end{figure}

The previous test presents a very simple geometry, but in real atrophysical problems the geometry of the medium can be quite complicated. With the purpose of studying the accuracy of the tree-based method in a highly structured medium, we analyze the case of a turbulent molecular cloud. This cloud has been formed through a simulation of colliding flows \citep{audit2005,vazquezsemadeni2007} with a turbulent velocity profile. This simulation produces a cloud that presents filaments and clumpy structures with densities ranging over more than six orders of magnitude. The size of the simulation box is $50$~pc, and we used two AMR levels ($\ell_{min}=7$ and $\ell_{max}=9$), with an equivalent maximal resolution of $512^3$ or, equivalently, a spatial resolution of $0.1$~pc. The criterion for refinement is again the density ($n = 50\ \mathrm{cm}^{-3}$ for the first refinement and $n = 100\ \mathrm{cm}^{-3}$ for the second one).

   \begin{figure}
   \centering
   \includegraphics[width=9.3cm]{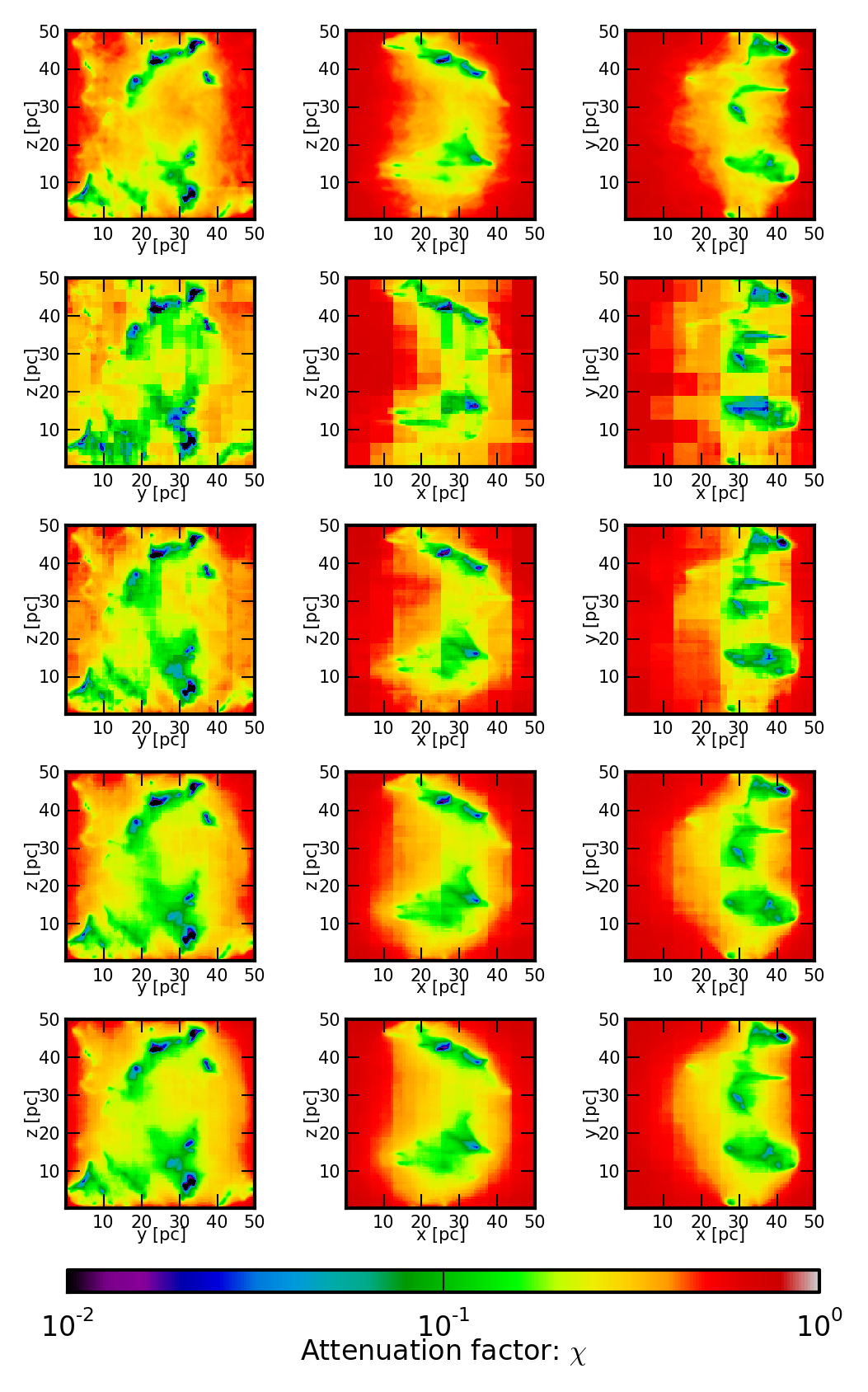}
   \caption{Extinction maps as seen by the cells in the midplanes. On the top we show the reference, calculated with the ray-tracing method. The other maps (from top to bottom) have been calculated with the tree-based method using $6$, $12$, $40$, and $84$ directions.}
   \label{Fig_cloudX}
   \end{figure}

   \begin{figure}
   \centering
   \includegraphics[width=9cm]{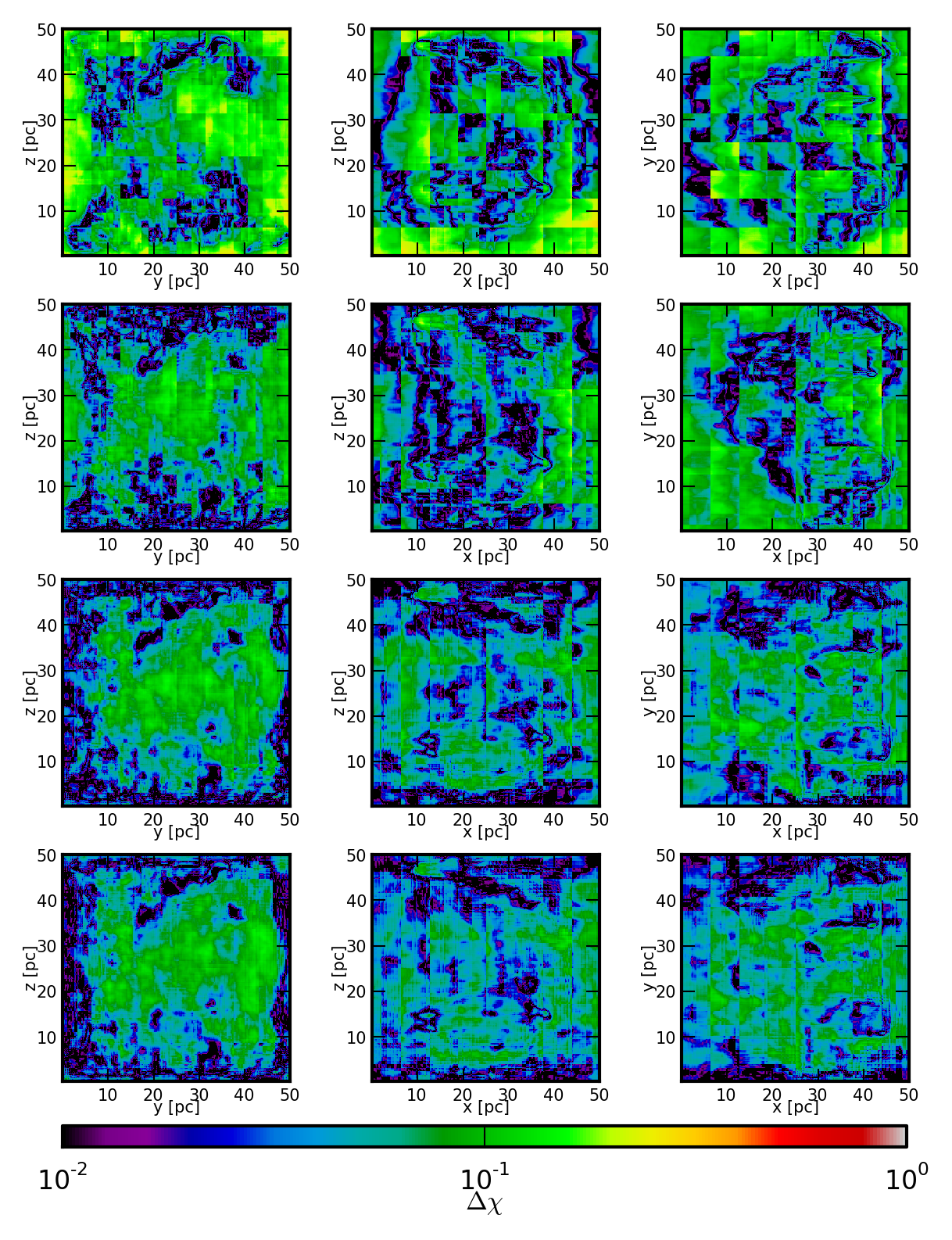}
   \caption{Relative difference maps for the tree-based method. From top to bottom: $6$, $12$, $40$, and $84$ directions.}
   \label{err_cloudX}
   \end{figure}

Figure~\ref{Fig_cloudN}  presents the column densities integrated along the main cartesian directions as seen by the cells in the midplanes. At the top we show the reference maps, calculated with the ray-tracing method and, in the middle, calculated with our tree-based method. At the bottom we present the relative error maps calculated according to Eq.~(\ref{eq1}). The mean value of the relative error is about $50 \%$, but for many cells it can reach up to more than $100\%$. The reference corresponds to an exact integration along the $x$, $y$, and $z$ axes, so even far structures are taken into account at the highest resolution. With the tree-based method, these far structures are distributed over a larger region. This can be seen by comparing the column density maps obtained with the tree-based method to the reference. The missing structures can be interpreted as structures far from the midplane, while the clumps seen using the tree-based method are structures close to the midplane.

For the turbulent cloud we present a $4\pi\ \mathrm{sr}$ map similar to Fig.~\ref{Fig_sph_theta_phi}. Figure \ref{Fig_cloud_theta_phi} shows the column density projection in every direction as seen by a cell sitting at the center of the turbulent cloud for $12$, $40$, $84$, $144$, and $220$ directions using the gather approach , the tree-based method, and  the associated relative error. This map, unlike the case for the spherical cloud, does not show such high errors. Moreover, the errors seem to be distributed evenly throughout the map. 

Figure~\ref{Fig_cloudX} presents the extinction maps. At the top we present the reference maps calculated with the exact ray-tracing method, and the rest of the maps correspond to the tree-based method. In Fig.~\ref{err_cloudX} we present the difference maps calculated as defined before. These figures show how the accuracy of the maps can be improved by including directions that are not aligned with the main cartesian directions, because more cells are taken into account.  The mean and maximum differences are summarized in Table \ref{table2}. When we use six directions, the mean difference is about $0.079$, while for the rest it is about $0.05$. Beyond the mean difference we see that the `six-ray' method presents large and systematic variations that are absent with the multiray approach.
As we have shown before for the spherical cloud test, the increasing number of directions does not improve the accuracy of the map considerably, particularly between $40$ and $84$ directions.

   \begin{figure}
   \centering
   \includegraphics[width=9cm]{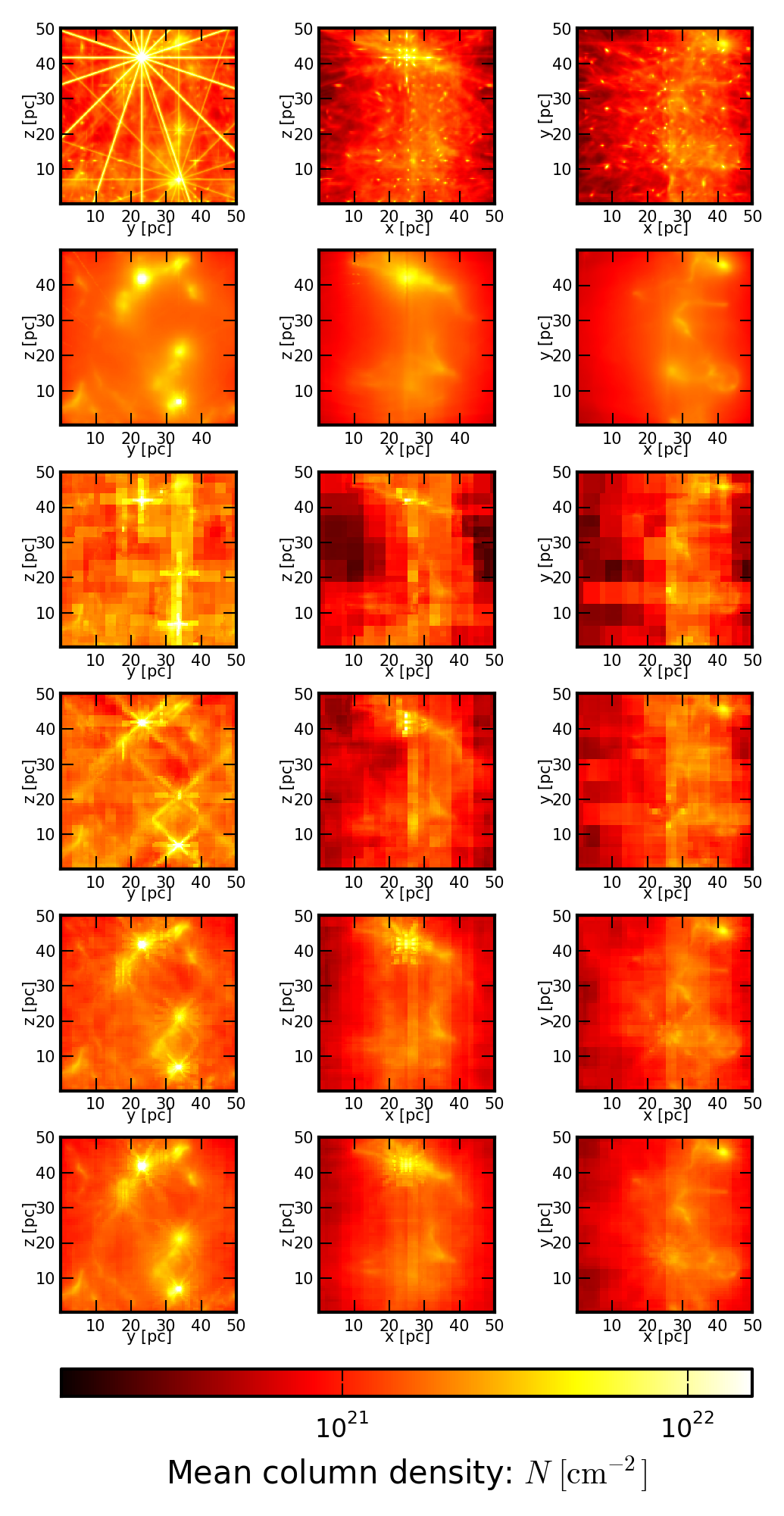} 
   \caption{Mean column density as seen by cells in the midplanes. From top to bottom: using the ray-tracing method for $50$ rays, using the gather approach for $220$ directions, and using the tree-based method for $6$, $12$, $40$, and $84$ directions.}
   \label{meancoldens}
   \end{figure}

In Fig.~\ref{meancoldens} we present the mean column density seen by cells in the midplanes. At the top we present the mean colum density maps done with the ray-tracing method. This method present strong shadowing effects for all the cells that are aligned with dense clumps, while for directions that are not aligned, the entire contribution is missed. For this reason we have included reference maps done with the gather method using $220$ angular bins (second row). The prohibitive cost of producing these maps led us to calculate the maps for the cuts at $x = 25 \ \mathrm{pc}$ and $y = 25 \ \mathrm{pc}$ at a lower resolution for a visual comparison. The map for the cut at $z = 25 \ \mathrm{pc}$ was calculated at level $\ell = 9$, and it was used to calculate the relative error. This figure shows how the estimation of the value of the mean column density improves as we increase the number of directions used in the tree-based method. In particular the mean error decreases as the number of directions increases, as shown in Table \ref{tablemean}.  
\begin{table}
\caption{Mean error for the mean column density maps.}
\label{tablemean} % is used to refer this table in the text
\centering
\begin{tabular}{l l}
\hline\hline
Directions			&  Mean error \\ % table heading
 $M\times N$		& $\langle \ |\langle N\rangle -\langle \bar N\rangle | /\langle\bar N\rangle \ \rangle$	\\
\hline \\
$6$				&	$0.278$		\\ 
$3 \times 4 = 12$	&	$0.211$		\\
$5 \times 8 = 40$	&	$0.135$		\\ 
$7 \times 12= 84$	&	$0.129$		\\ 
\hline
\end{tabular}
\end{table}

%DISCUSSION, JUSTIFICATION ON THE ACCURACY, 
 \subsection{General remarks}\label{remarks}

That the accuracy does not improve considerably is consistent with the results of \cite{clark2011} for  TreeCol, where they show that the efficiency depends on the relative sizes of the opening angle and the angular size of the node. This is true for a pure gather approach. It is important to notice that our implementation of the tree-based method is a hybrid method. The decreasing resolution for far cells mimics a gather approach, but calculation of the contributions to the column densities is done as in a ray-tracing method. With this in mind, we note that the increasing number of directions does not change the angular resolution for a given cell, but it does improve the quality of the description of the density field for calculating the extinction. This can be seen in Fig.~\ref{Fig_cloudX}, where the extinction maps are smoother and the maximum error is reduced as the number of directions is increased. Then the optimal number of directions for the tree-based method will correspond to the best compromise between resolution and numerical cost. Taking this into account and using as a criterium the variation of the error, a good compromise is found for $40$ directions.

The difference maps for the extinction shown in Figs.~\ref{err_sphereX} and \ref{err_cloudX} present very different features. In particular, the highest error happens toward the border of the computational domain for the spherical case, while the error in this region is much smaller for the turbulent cloud. This is probably due to the different distribution of sources. This effect is more evident for the spherical case than for the turbulent cloud owing to the anisotropy of the density field. For the turbulent cloud the distribution of sources is more isotropic, and for each cell it will be easier to cast a source, while for a compact central distribution, especially for cells not aligned with the central source, it will be more difficult to describe the density field. This same effect can be seen in Figs.~\ref{Fig_cloud_theta_phi} and \ref{Fig_sph_theta_phi}.      

%%%%%%%%%%%%%%%%%%%%%%%%% 	PREC TEST	    %%%%%%%%%%%%%%%%%%%%%%%%%%%%%%
\subsection{Validation of the precalculation approach}\label{validationprec}

%The aim of this section is to show how much faster the code gets when the screening is used, and how the two-level approximation used in order to speed up the code affects the accuracy of the extinction maps. 
To quantify the influence on the accuracy of estimating the attenuation factor of the two-level approximation and the influence of precalculating geometrical terms on the speed of the code, we compared both implementations of the tree-based method. The first implementation calculates each contribution to the column density for each cell without doing any approximation. This implementation is called \emph{in situ}. The second implementation uses the precalculation of geometrical terms and the two-level approximation. We present a comparison of the column density maps integrated along three different directions. For the extinction we present a comparison of the extinction maps and the difference on the estimated value of the extinction. We also present the calculation CPU times ($t_{sim}$) relative to the standard CPU time ($t_0$). The standard CPU time is the calculation time of a simulation that does not include the extinction. The two-level approximation accelerates the code, but it shifts the point of reference that defines the directions (Fig.~\ref{2levelcorrection}) introducing errors on the calculation. 

\begin{table}
\caption{Calculation time relative to the case without screening and the average difference of extinction maps.}
\label{table1} 
\centering
\begin{tabular}{l l l l }
\hline\hline
Directions			& \multicolumn{2}{c}{$t_{sim}/t_{0}$}	&	$\langle|\chi_{prec} - \chi_{in\ situ}|\rangle$\\ % table heading
 $M\times N$		& {in situ}		&  {prec} 	&	\\
\hline \\
$6$				&	$1.1$	&	\textendash	&	\textendash\ \\
$3 \times 4 = 12$	&	$7.9$	&	$\textbf{1.6}$	&	$0.0028$	\\
$5 \times 8 = 40$	&	$15.8$	&	$\textbf{2.4}$	&	$0.0037$\\
$7 \times 12= 84$	&	$26.3$	&	$\textbf{3.3}$	&	$0.0034$\\
\hline
\end{tabular}

\end{table}

  \begin{figure}
   \centering
   \includegraphics[width=9cm]{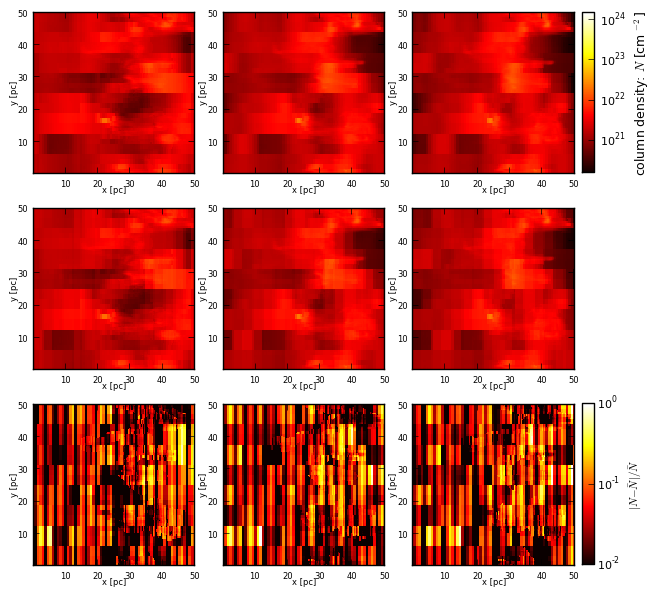}
   \caption{Influence of the precalculation module. Column density maps for three directions not aligned with the cartesian directions for the case of calculation \emph{in situ} (top), using the precalculation module (middle), and the relative difference map (bottom). The azimuthal angle is fixed at $\phi = 0$, and from left to right the polar angle is $\theta = 0.841, 0.644, 0.541 \ \radian$. The mean fractional difference, defined as in Eq.~(\ref{eq1}), are $\Delta N = 6.3, 6.6, 6.8 \%$, respectively.}
    \label{Nnoprec_prec}
    \end{figure}

  \begin{figure}
   \centering
   \includegraphics[width=9cm]{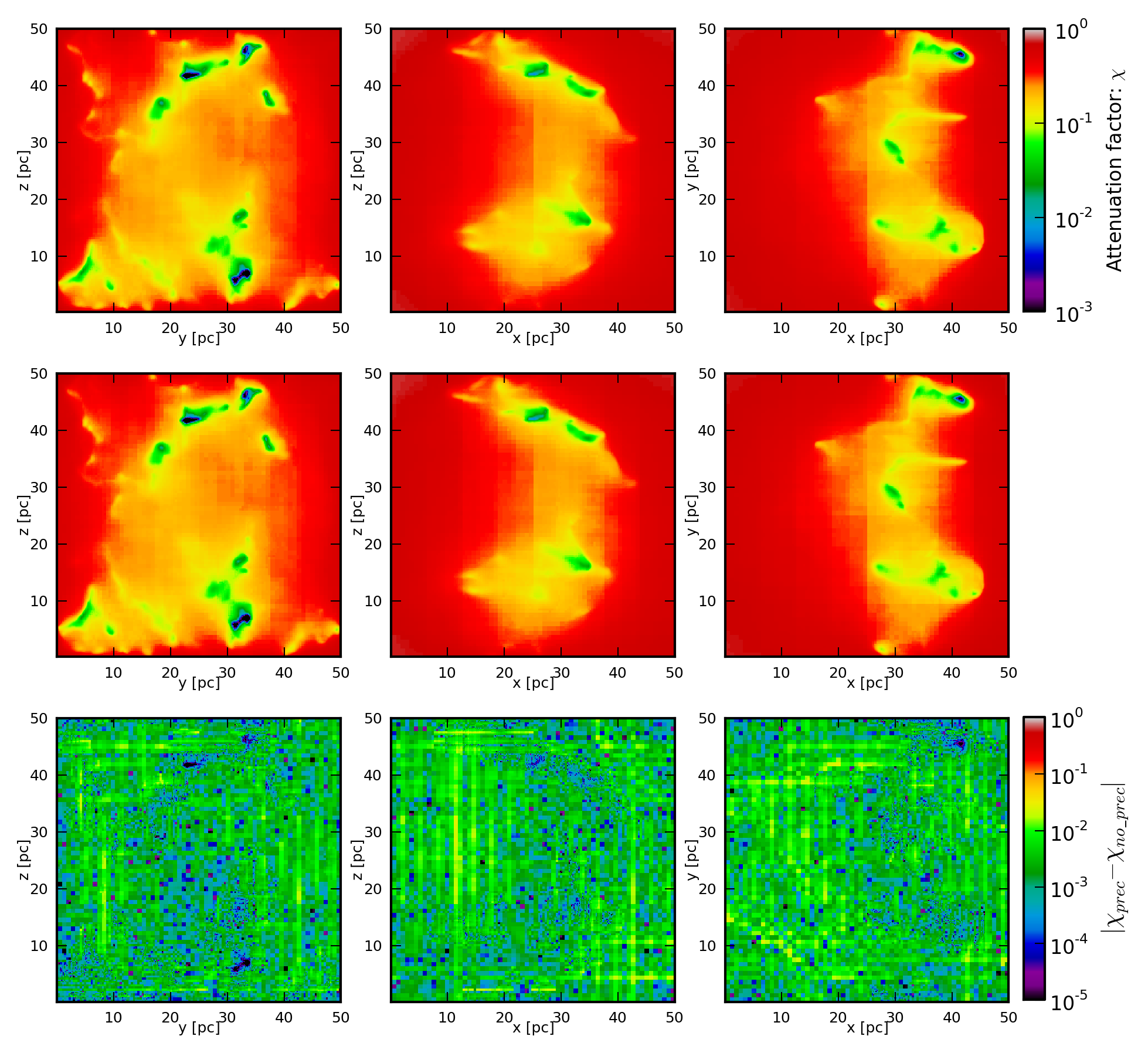}
   \caption{Influence of the precalculation module. Extinction maps calculated using $84$ directions ($M=7$, $N=12$). At the top we present the extinction map for the case where all the calculations are done \emph{in situ}, and in the middle using the precalculation module. At the bottom we present difference maps.}
    \label{noprec_prec}
    \end{figure}

To estimate the gain in performance and the induced errors, we performed eight different calculations. The first one corresponds to the reference, calculated without including the screening for the UV. The second one uses a simple implementation of our method using just 
six directions, and it does not need any kind of geometrical correction. The six other simulations use two implementations of our method (calculation \emph{in situ} and precalculation module) for $3\times 4=12$, $5\times 8=40$, and $7\times 12=84$ directions. We use a turbulent cloud as initial condition, which is created by converging flows without any screening, and we compare the calculation time. For the comparison of the column density maps and the extinction maps we use  the initial output, in order to use identical clouds. Table \ref{table1} presents the simulation times relative to $t_0$, the calculation time without screening, and the mean value of the difference in the extinction maps. Calculating column densities for the case using six directions does not require geometrical corrections, and the calculation time is very close to the reference case. This means that the geometrical corrections are expensive. The precalculation module significantly reduces the calculation time, and the code can be up to eight times faster when the precalculation module is used. 

For estimating of the induced differences related to the two-level correction we calculated  column density maps and extinction maps for both implementations.
For the column density maps, we selected three directions that are not aligned with any of the main cartesian directions $\pm x, \pm y, \pm z$, because these directions are not sensitive to this approximation. Figure~\ref{Nnoprec_prec} shows the column density maps for three directions where we fixed the azimuthal angle ($\phi =0$) and we varied the polar angle ($\theta = 0.841, 0.644, 0.541 \ \radian$). The relative difference map was calculated according to Eq.~(\ref{eq1}), and the mean value for all of them is less than $7\%$.

For the extinction we present a comparison between both implementations. We calculated the difference of the extinction maps as seen by the cells in the midplanes according to Eq.~(\ref{eq2}), for the case where we introduce the precalculation module with respect to the case where all the calculations are done \emph{in situ}. This comparison was done using $12$, $40$, and $84$ directions. Table \ref{table1} presents the mean values of the difference, which are better than $0.004$ for all the cases, and it does not depend on the number of directions used. As an example, we present the extinction maps  and difference maps for the cells in the midplanes in Fig.~\ref{noprec_prec} as a comparison of both implementations for the case where we use $84$ directions ($7$ intervals for the polar angle and $12$ for the azimuthal angle). Overall we conclude that the method with precalculation is sufficiently accurate and considerably faster than the \emph{in situ} method. 

%%%%%%%%%%%%%%%%%%%%%%%%%%%%%%%%%%%%%%%%%%%%%%%%%%%%%%%%%%%%%%%%%%

\section{Application}  \label{application} %%% SCIENCE %%%

%%%%%%%%%%%%%%%%%%%%%%%%
% SLICE, General structure of the cloud
   \begin{figure*}
   \centering
   \includegraphics[width=14cm]{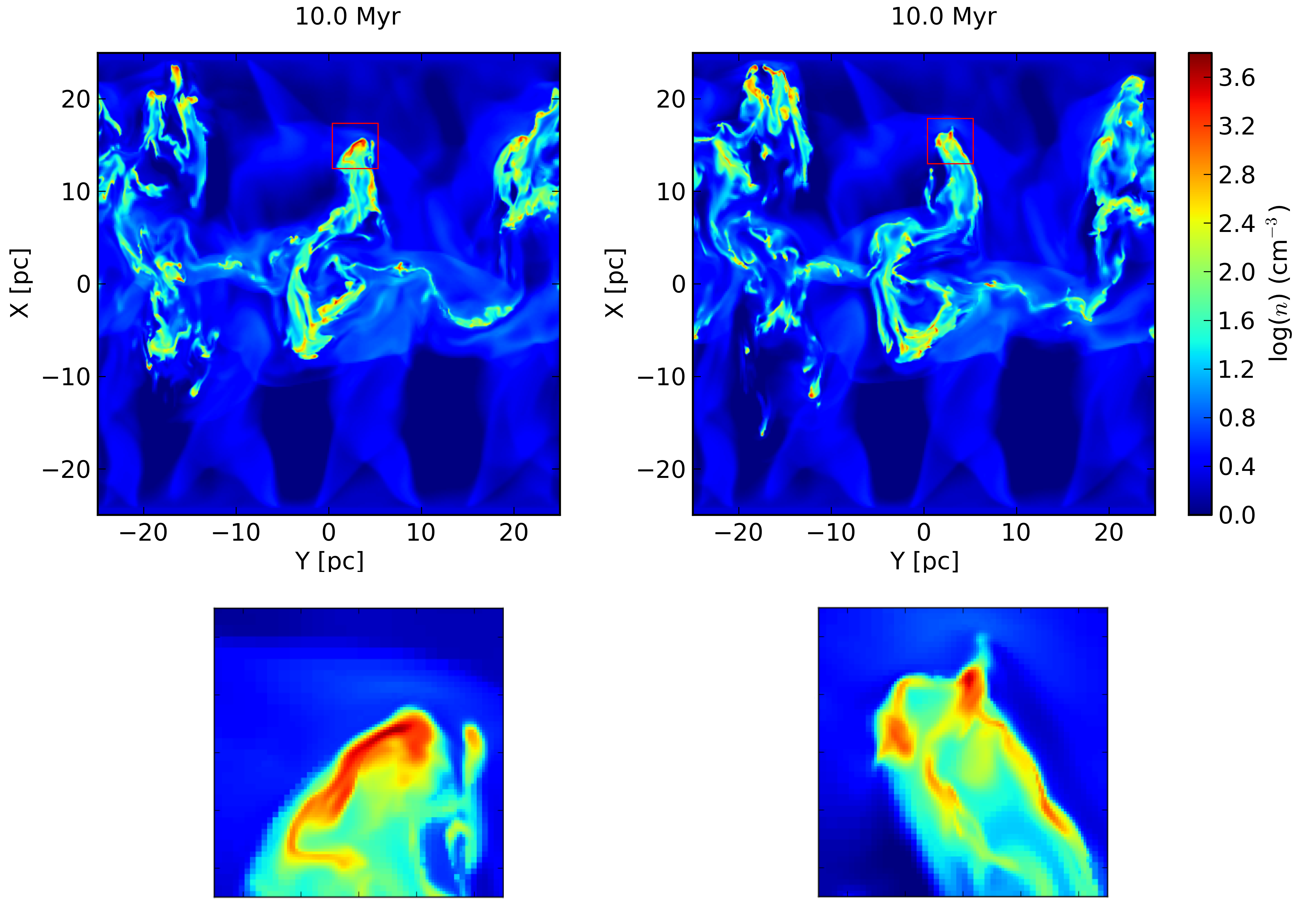}
   \caption{Slices cut through the midplane at $t=10$ Myr showing the detail of difference in fragmentation. The left panel shows the case without screening, and the right panel shows the case with screening. The fragments zoomed in have a size of $5$~pc and are centered at $x=15,\ y=2.75$~pc with respect to the center of the box.}
   \label{slice}
   \end{figure*}
%%%%%%%%%%%%%%%%%%%%%%%%

As an application we study the formation and the evolution of a molecular cloud formed from colliding streams of warm atomic gas \citep{audit2005,vazquezsemadeni2007, heitsch_et_al_2005, heitsch_et_al_2006, heitsch_et_al_2008}. The 3D simulations were performed using the AMR code RAMSES. 

The set-up and the initial conditions are similar to those of \cite{hennebelleetal2008} \citep[see also][]{vazquezsemadeni2007, heitsch_et_al_2005, heitsch_et_al_2006, heitsch_et_al_2008, inoue2012}. We consider a cubic box of length $L=50$~pc. We allow two AMR levels, with $\ell_{min} = 8$ and $\ell_{max} = 10$, reaching an effective numerical resolution of $1024^3$ cells and a spatial resolution of about $0.05$~pc. The boundary conditions are imposed to mimic the large scale converging flows. The gas is injected from the left and right faces of the simulation box with a weakly turbulent velocity and a density of $2\  \mathrm{cm}^{-3}$ at a temperature of $8000\ \kelvin$. For the remaining faces, we use periodic boundary conditions. The velocity field of the incoming gas $V_{in}$ depends on $y$, with an average velocity $V_0 = 15\ \mathrm{km\ s^{-1}}$ and modulated by a function of amplitude $\epsilon = 0.5$, as defined in \cite{audit2005}. Initially the simulation box is filled with warm atomic gas with the same density and temperature as the inflowing gas and is uniformly magnetized with a magnetic field of strength $2.5\ \micro$G parallel to the $x$ axis and therefore aligned with the incoming velocity field. The gas in the simulation box is heated by the background UV field, which corresponds to the Draine field $G_0 = 1.7$ in Habing units. Because this radiation field is assumed to be isotropic and constant, then we can define the effective UV field as
\begin{equation}
	\overline{G}_0 = \chi G_0 		\label{G0_eff}
\end{equation}

\noindent where $\chi$ is the attenuation factor for the UV flux defined in Eq.~(\ref{eq_chi}) and calculated for each \emph{leaf cell} in the simulation. 

We use the same cooling and heating functions than \cite{audit2005}, but we have modified the heating by adding the screening for the UV and we included the heating by cosmic rays. The heating rates as implemented in the code are

%We use the  heating rate, but modified in order to use the effective incident UV field $\overline G_0$, and the :  
\begin{eqnarray}
	\Gamma_{UV} &=& 1.0\times10^{-24}\ \varepsilon\ \ \overline G_0\ n\ \mathrm{ergs\ cm^{-3} s^{-1}} \label{eq_gamma_uv}\\
	\Gamma_{CR} &=& 10^{-27}\ n\ \mathrm{ergs\ cm^{-3} s^{-1}} \label{eq_gamma_cr}
\end{eqnarray}

\noindent where Eq.~(\ref{eq_gamma_uv}) is the heating rate due to the photoelectric effect on small grains and PAHs due to the FUV radiation \citep{bakestielens1994}, $n$ is the hydrogen density in $\mathrm{cm^{-3}}$,  $\varepsilon$ is the heating efficiency, calculated as in \cite{wolfire1995} : \\$
	\varepsilon =  \frac{4.9\times 10^{-2}}{1.0 + \left [ \left( \overline G_0 T^{1/2}/n_e \right) /1925 \right]^{0.73}} + \frac{3.7\times 10^{-2} \left(T/10^{4}\right)^{0.7}}{1.0 + \left [ \left( \overline G_0 T^{1/2}/n_e \right) /5000 \right]}$, \\
\noindent with $n_e$, the electron density, given by the approximation proposed by \cite{wolfire2003}. Equation~(\ref{eq_gamma_cr}) is the heating rate due to cosmic rays according to the intermediate value given by \cite{goldsmith2001}.

Our aim is to understand how the screening caused by the surrounding matter can affect the gas distribution and the distribution of structures formed by diminishing the amount of radiation that arrives at the cells. Then we present two simulations in order to compare the influence of the screening for the UV. The first simulation does not include this effect, which is equivalent to considering the gas as optically thin and having an attenuation factor $\chi = 1$. This means that the radiation reaches the cells unchanged. On the other hand, the second simulation takes the absorption due to the surrounding material into account. In this case the estimation is done using the tree-based method, as described in section \ref{sec_method}. Just for the sake of simplicity and time, we used $12$ directions ($M=3$ intervals for the polar angle and $N=4$ for the azimuthal angle) for calculating the attenuation factor for the UV field (but see the Appendix~\ref{app_lowres} for a comparison between simulations at lower resolution using different number of directions to calculate the extinction for the UV field). 

\subsection{General structure of the cloud}

To understand how the screening for the UV affects the structure of molecular clouds, we present a comparison between both simulations in Fig.~\ref{slice} by presenting the local density in a slice cut through the middle of the simulation box at  $t=10$ Myr. This figure shows that the large scale structure of the molecular cloud seems to be barely affected by the UV screening, but the detail of the fragmentation of dense structures is substantially influenced by the extinction. In the case where the extinction is not taken into account, the structures formed tend to be bigger, while when the screening is included, the same region seems to be torn up, presenting smaller structures. This seems to indicate that the extinction mainly affects the dense parts of the gas. To go beyond this qualitative impression, we now turn to more quantitative studies.

\subsection{Probability distribution function}

\begin{figure}
   \centering
   \includegraphics[width=9cm]{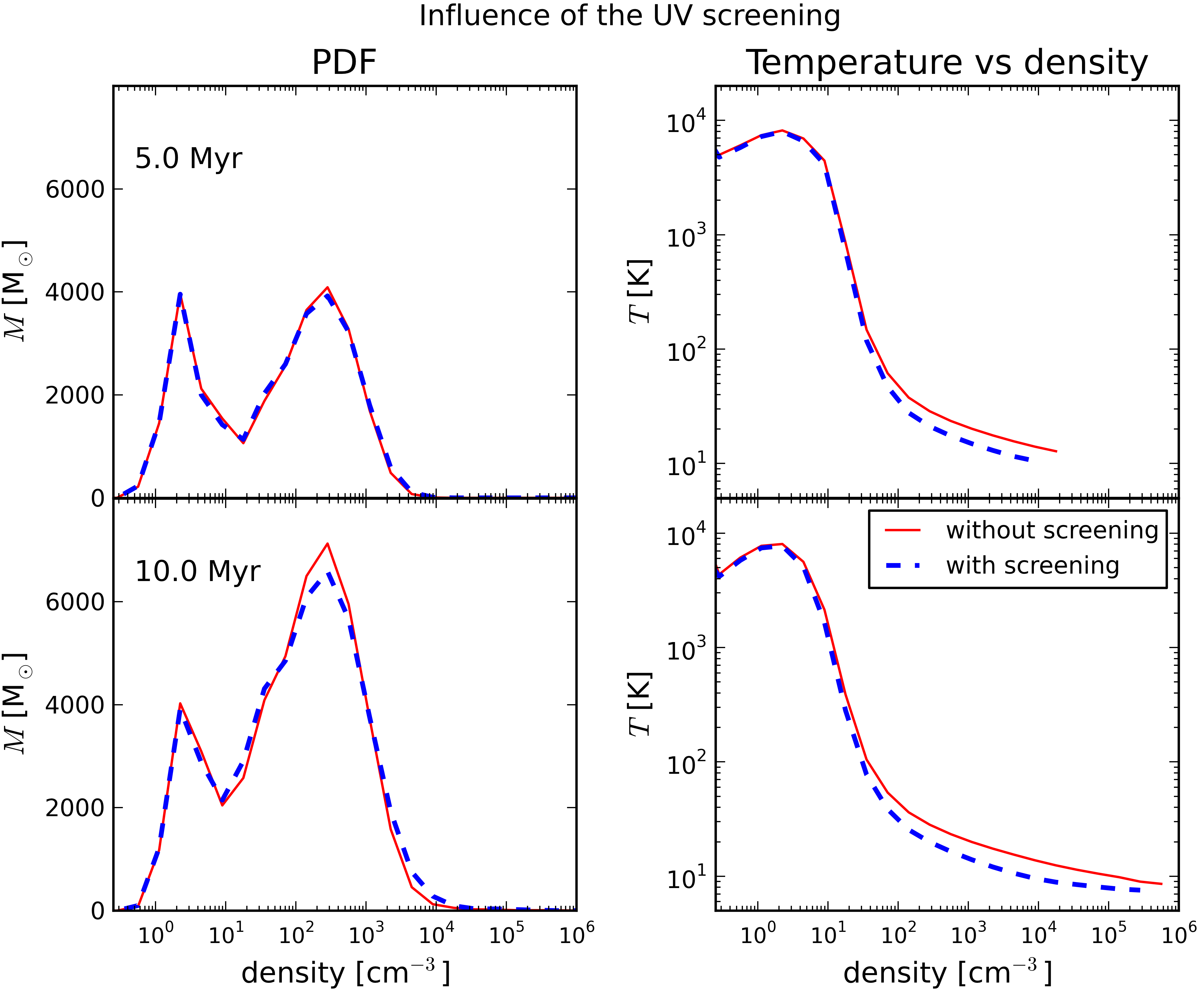}
   \caption{Comparison of the probability distribution  function (PDF) of gas (left panel) and the corresponding temperature per density bin (right panel) at $t = 5$ Myr (top) and at $t = 10$ Myr (bottom) for the cases with and without screening. }
   \label{pdf}
   \end{figure}

The density probability distribution function (PDF) corresponds to the distribution of mass as a function of density and is one of the simplest statistical tools for understanding how the gas is distributed. In Fig.~\ref{pdf} we present a comparison of the PDFs and the mass-averaged temperature for each density bin. For both cases we can see that starting with only warm gas, which is caused by the cooling processes present in the simulation, the gas is able to transit from the WNM phase to the CNM phase \citep{audit2005}. As the gas enters the box, the CNM phase develops.  The PDF of the gas density and the temperature distribution show the bistable nature of the medium \citep{field1969, wolfire1995, wolfire2003}. The panel on the right shows how the screening affects mainly the dense medium, while the warm gas is almost not altered. Initially the distribution is not significantly affected. However the difference can be seen as dense gas forms. The bottom left hand panel shows that a part of the gas in the CNM phase presents higher density and lower temperature. Consequently, part of the gas has a lower density owing to the rarefaction caused by the development of more compact structures, and this explains why there is more gas with $n\sim 300 ~ \mathrm{cm}^{-3}$ in the case without screening.

%%%%%%%%%%%%%%%%%%%%%%%%%
% PDF T vs n
%%%%%%%%%%%%%%%%%%%%%%%%%

\subsection{Mass spectra}

\begin{figure*}
   \centering
   \includegraphics[width=17cm]{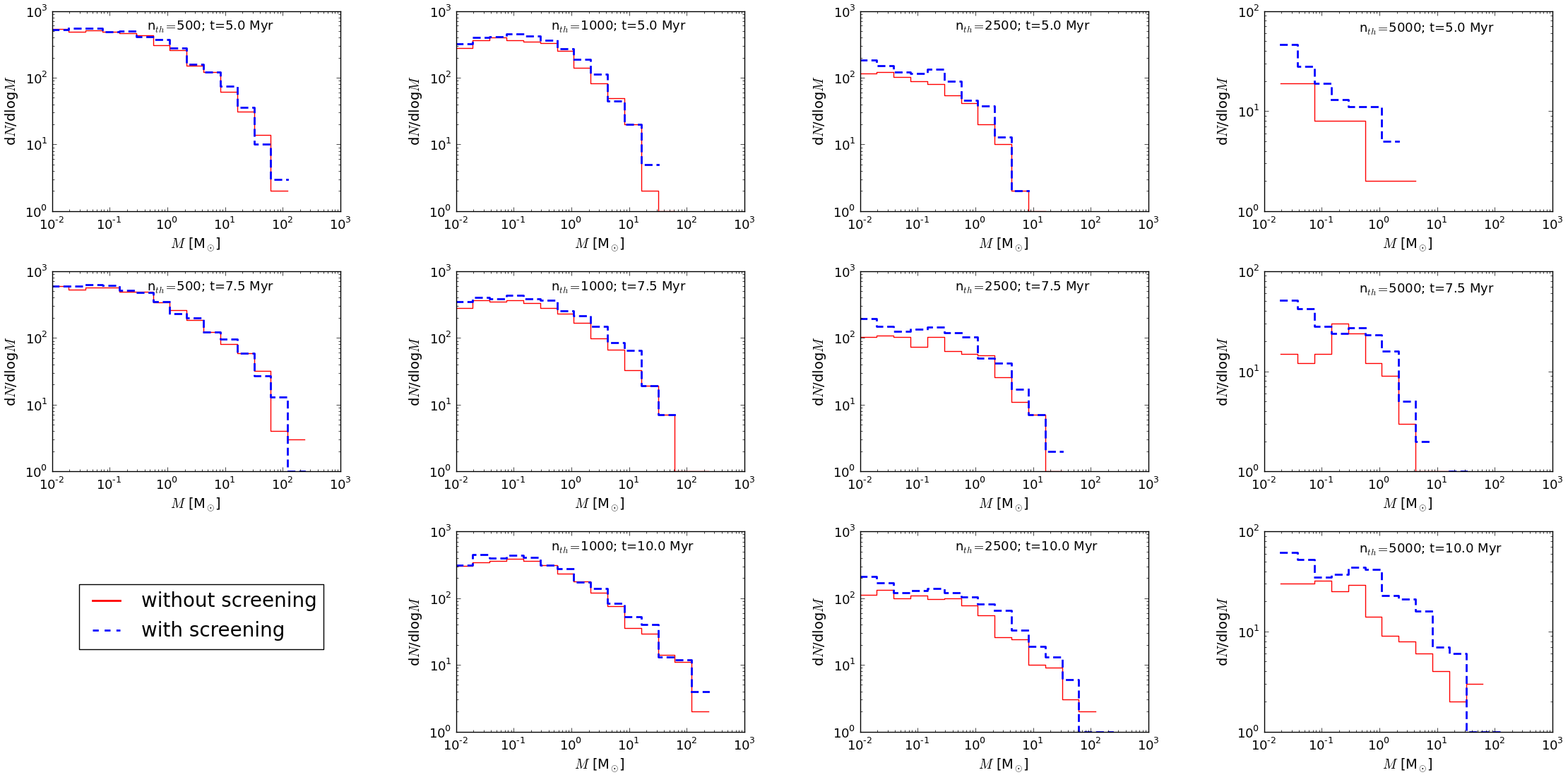}
   \caption{Comparison of the mass spectra. From top to bottom at $t = 5, 7.5, 10$ Myr. From left to right for the density thresholds: $n_{th}~=~500, 1000, 2500, 5000~\mathrm{cm^{-3}}$.  }
   \label{mspec}
   \end{figure*}

To investigate the influence of the UV screening on the statistical properties of the structures formed within the clouds, we analyzed the mass spectra of the clumps for both cases. The mass spectrum presents the number of structures per logarithmic mass interval for a given density threshold. For the clump extraction, we selected all the cells with a density higher than a given density threshold $n_{th}$, and using a friend-of-friend algorithm, we identified the spatially connected regions that constitute a clump, rejecting isolated cells. In Fig.~\ref{mspec} we present the evolution of the clump distribution for $t = 5,\ 7.5$, and $10$ Myr for different density thresholds. We show that the mass spectrum for low density thresholds does not vary significantly when we include the UV screening. On the other hand, the differences in the mass spectrum are more pronounced for more compact structures ($n_{th} \ge 2500\ \mathrm{cm^{-1}}$). The number of compact clumps found is systematically larger in the simulation that includes the UV screening. 

%%%%%%%%%%%%%%%%%%%%
%MASS SPEC
%%%%%%%%%%%%%%%%%%

\section{Conclusions} \label{conclusions}

%Method conclusions\\

We have introduced a tree-based method for a fast estimation of column densities in astrophysical simulations. The general idea of this method can be implemented in any code with a tree-based data structure, and the implementation strategy is not unique. We presented a simple implementation on the AMR code RAMSES. In particular, we used a precalculation module that speeds up the calculation considerably without changing the accuracy of the estimation significantly (better than $7\%$ for the column density maps and about $0.3 \%$ for the extinction maps). 

Since the cells that contribute to the column density are taken into account at lower resolution as the distance increases, and if the tree is fully threaded, the tree-based method only needs the information stored locally in the essential tree and does not need any communication between different CPUs, making the method suitable for parallel computing. Finally, the errors on the column density maps are generally about $50 \%$, while the extinction maps calculated from the estimated column densities have errors lower than $10\%$.

%Scientific conlusions: Warm gas\\

We found that the UV screening does not have a strong influence on the general structure of molecular clouds, but it has a significant impact on the details of fragments. Most notably, the extinction for the UV mainly affects the dense gas. The temperature for the WNM phase remains almost unchanged, while for the CNM it is lower by a factor up to $50 \%$ when the shielding effect of dust is included. Consequently, the local value of the Jeans mass is reduced, favoring the gravitational collapse of smaller structures. This is evident in the comparison of the mass spectra, where for low-density thresholds there are no noticeable differences; however, the number of compact structures for higher density thresholds is greater for the case that includes the screening. 

Because an important part of the chemistry in the ISM depends on the UV radiation and on temperature, this method can be applied to the interstellar chemistry to more realistically estimate these parameters. Estimates of column densities can also be used to give a better value for the CR ionization rate.

\begin{acknowledgements}
V.V. acknowledges support from a CNRS-CONICYT scholarship. This research has been partially funded by CONICYT and CNRS, according to the December 11, 2007 agreement.\\
P.H. acknowledges the finantial support of the Agence National pour la Recherche through the COSMIS project. This research has received funding from the European Research Council under the European Community’s Seventh Framework Program (FP7/2007-2013 Grant Agreement No. 306483).

\end{acknowledgements}

\bibliographystyle{bibtex/aa} % style aa.bst 
\bibliography{biblio_val} % your references Yourfile.bib

%%%%%%%%%%%%%%%%%%%%%%%%%%%%%%%%%%%

\appendix

\section{Low resolution study}\label{app_lowres}

\begin{figure}
   \centering
   \includegraphics[width=8cm]{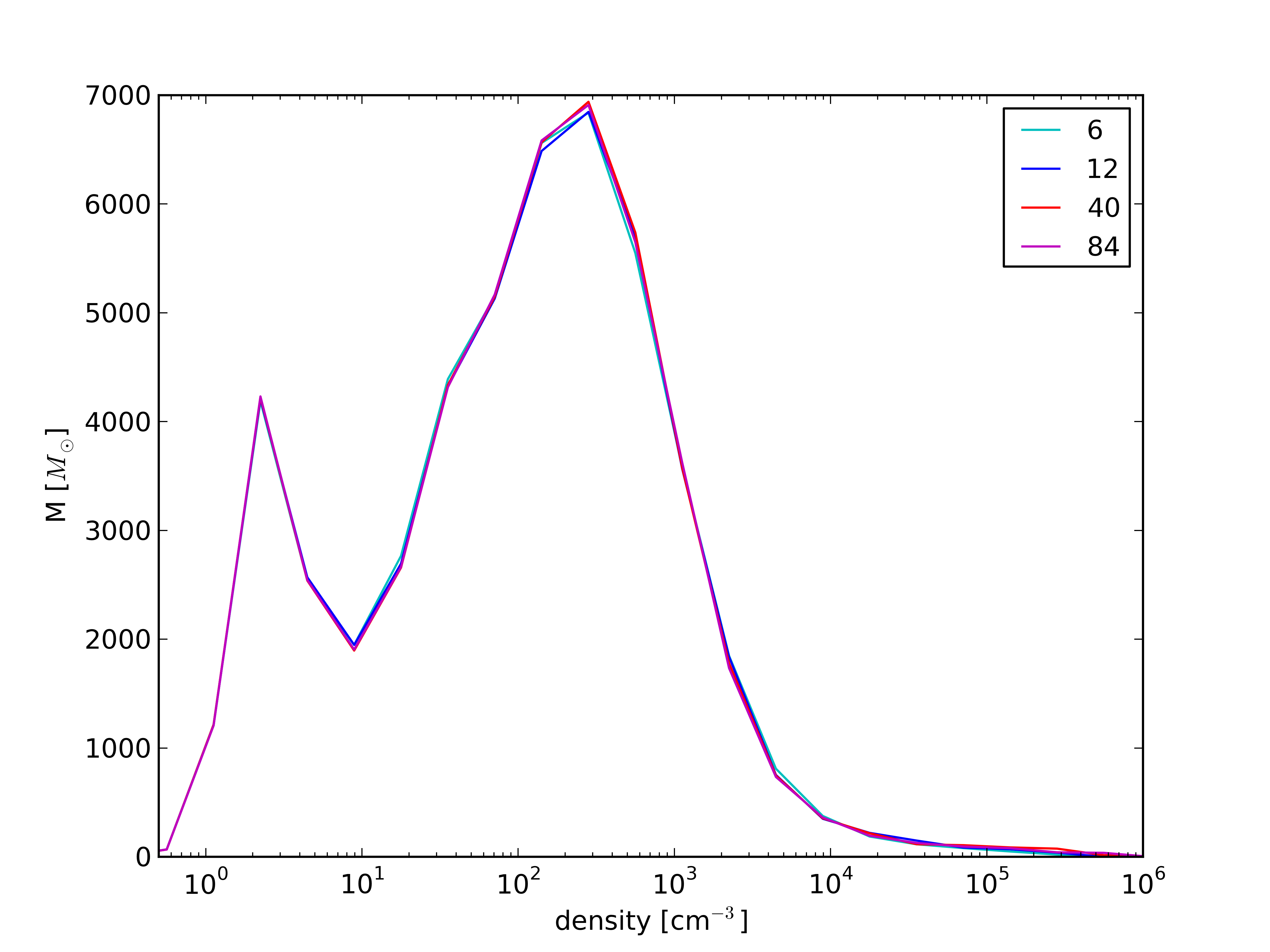} \\
   \includegraphics[width=8cm]{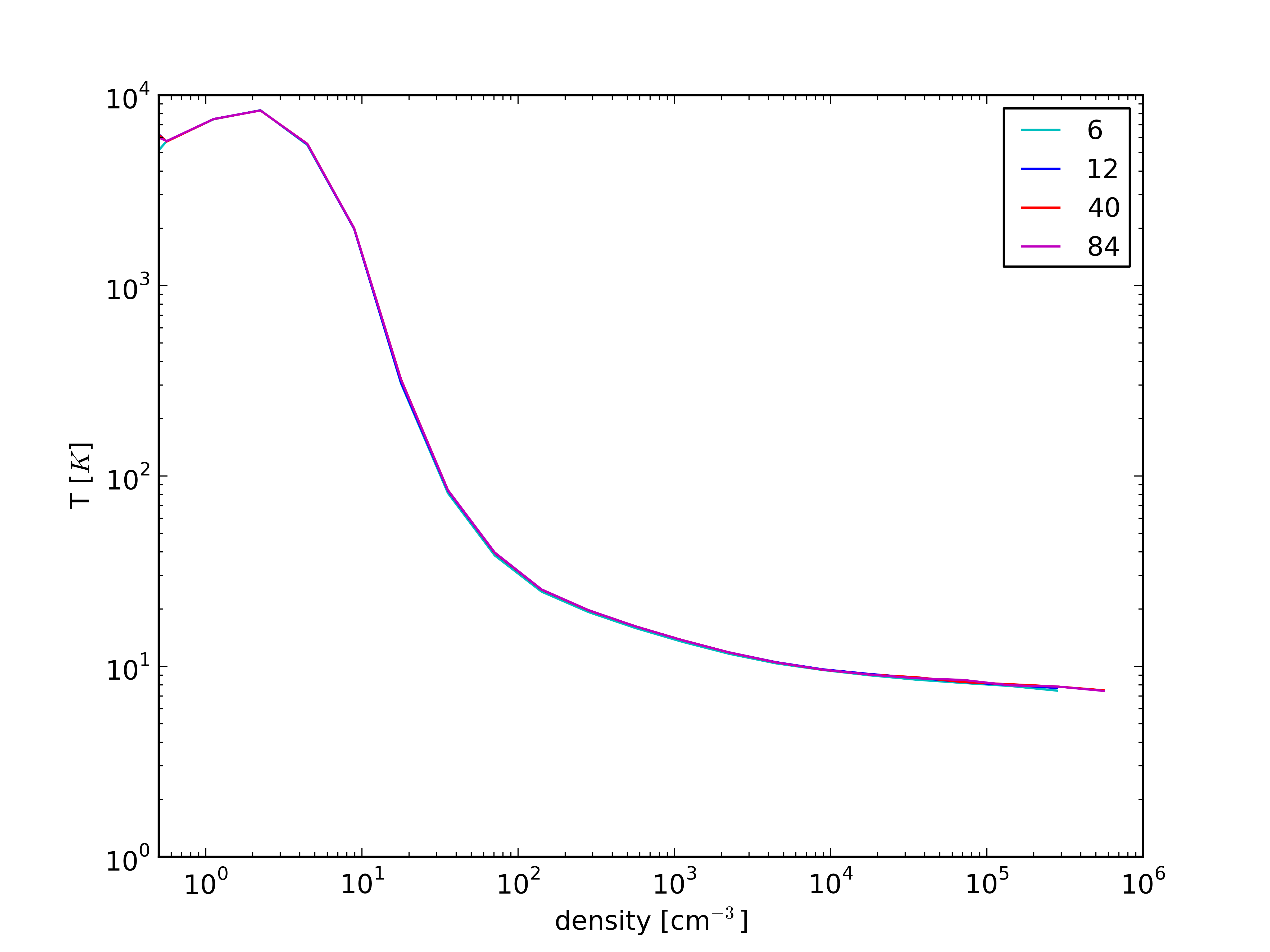}
   \caption{Comparison of the PDF (top) and the respective temperature per density bin (bottom) at $t = 10$ Myr, obtained including the screening calculated with the tree-based method using $6$, $12$, $40$, and $84$ directions.}
   \label{lowres_temp}
   \end{figure}

\begin{figure}
   \centering
   \includegraphics[width=7cm]{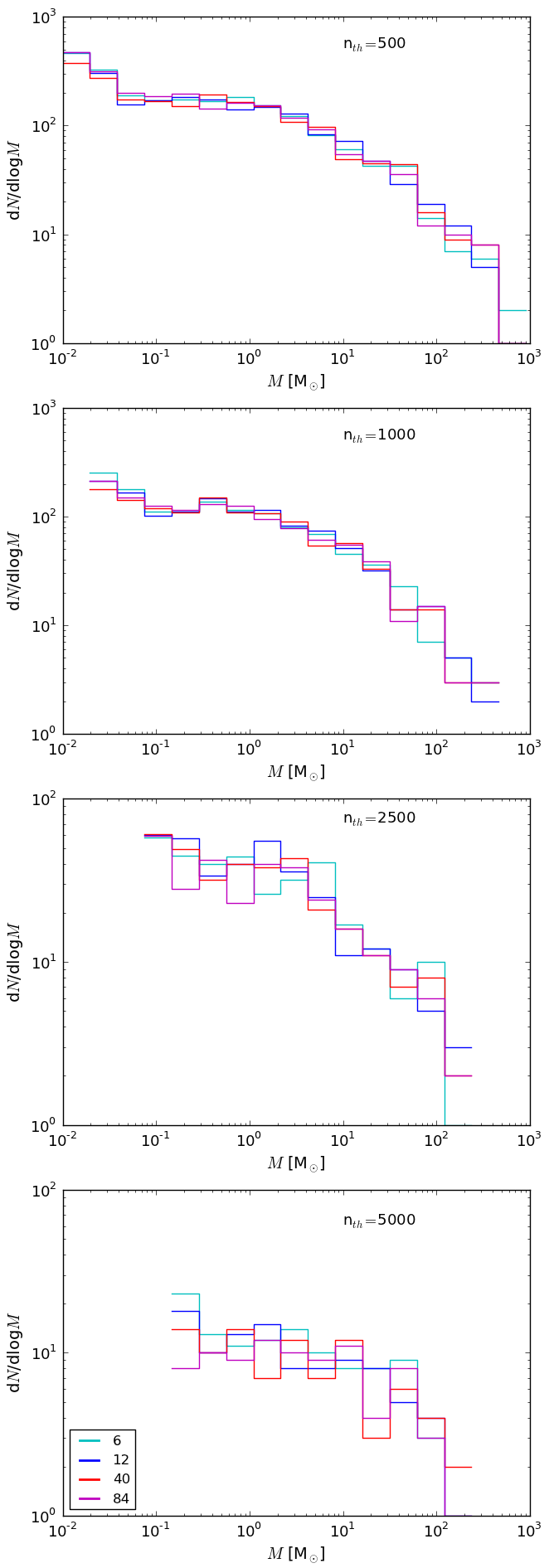}
   \caption{Comparison of the mass spectrum of clumps at $t = 10$~Myr for different estimations of the column density at low resolution. From top to bottom: $n_{th} = 500,\ 1000,\ 2500, 5000\ \mathrm{cm^{-3}}$.}
   \label{lowres_mspec}
   \end{figure}

%REWRITE THIS $\rightarrow$ ANNEX   
In order to ensure that the difference on the statistical properties depends on the inclusion of the screening but not on the method used for estimating it, we present a low resolution study of the influence of the method. We have performed four identical simulations, as decribed in section \ref{application}, where the screening has been calculated using $6$, $12$, $40$, and $84$ directions. We allow two AMR levels ($\ell_{min} = 7$ and $\ell_{max}=9$), with an effective numerical resolution of $512^3$ cells and spatial resolution of $0.1$~pc. We let them evolve until $t=10$ Myr. Figure~\ref{lowres_temp} presents a comparison of the probability distribution function (PDF) and respective temperature per density bin. This figure shows that the number of directions used to calculate the extinction is not crucial for the gas distribution and the temperature per density bin is almost not sensitive to the number of directions used. Figure~\ref{lowres_mspec} presents the comparison of the mass spectrum for different density thresolds. It indicates that the clump distribution does not change considerably with the number of directions and that the differences remain within the limits of the statistics.

%%%%%%%%%%%%%%%%%%%%%%%%%%%%%%%%%%%

\end{document}